\documentstyle[twoside,fleqn,epsfig]{article}

\def\be{\begin{equation}}
\def\ee{\end{equation}}
\def\bea{\begin{eqnarray}}
\def\eea{\end{eqnarray}}
\newcommand{\lsim}{\mathrel{\mathop{\kern 0pt \rlap
  {\raise.2ex\hbox{$<$}}} \lower.9ex\hbox{\kern-.190em $\sim$}}}
\newcommand{\gsim}{\mathrel{\mathop{\kern 0pt \rlap
  {\raise.2ex\hbox{$>$}}}
  \lower.9ex\hbox{\kern-.190em $\sim$}}}

\newcommand{\AmS}{{\protect\the\textfont2
  A\kern-.1667em\lower.5ex\hbox{M}\kern-.125emS}}
\normalsize
\begin{document}

\baselineskip=0.65cm

\begin{center}
\Large
{\bf Final model independent result of DAMA/LIBRA--phase1}
\rm
\end{center}

\large

\begin{center}

R.\,Bernabei$^{a,b}$,~P.\,Belli$^{b}$,~F.\,Cappella$^{c,d}$,~V.\,Caracciolo$^{e}$,~S.\,Castellano$^{e}$, 
\vspace{1mm}

R.\,Cerulli$^{e}$,~C.J.\,Dai$^{f}$,~A.\,d'Angelo$^{c,d}$,~S.\,d'Angelo$^{a,b}$,
\vspace{1mm}

~A. Di Marco$^{a,b}$,~H.L.\,He$^{f}$, A.\,Incicchitti$^{d}$,~H.H.\,Kuang$^{f}$,
\vspace{1mm}

~X.H.\,Ma$^{f}$,~F.\,Montecchia$^{b,g}$,~D.\,Prosperi$^{c,d,}\footnote{Deceased}$,
\vspace{1mm}

~X.D.\,Sheng$^{f}$,~R.G.\,Wang$^{f}$,~Z.P.\,Ye$^{f,h}$
\vspace{1mm}

\normalsize
\vspace{0.4cm}

$^{a}${\it Dip. di Fisica, Universit\`a di Roma ``Tor Vergata'', I-00133  Rome, Italy}
\vspace{1mm}

$^{b}${\it INFN, sez. Roma ``Tor Vergata'', I-00133 Rome, Italy}
\vspace{1mm}

$^{c}${\it Dip. di Fisica, Universit\`a di Roma ``La Sapienza'', I-00185 Rome, Italy}
\vspace{1mm}

$^{d}${\it INFN, sez. Roma, I-00185 Rome, Italy}
\vspace{1mm}

$^{e}${\it Laboratori Nazionali del Gran Sasso, I.N.F.N., Assergi, Italy}
\vspace{1mm}

$^{f}${\it IHEP, Chinese Academy, P.O. Box 918/3, Beijing 100039, China}
\vspace{1mm}

$^{g}${\it Dipartimento di Ingegneria Civile e Ingegneria Informatica, Universit\`a
di Roma ``Tor Vergata''}
\vspace{1mm}

$^{h}${\it University of Jing Gangshan, Jiangxi, China}

\end{center}
	
\normalsize

\begin{abstract}

The results obtained with the total exposure of 1.04 ton $\times$ yr collected by DAMA/LIBRA--phase1
deep underground at the Gran Sasso National Laboratory (LNGS) of the I.N.F.N. during 7 annual cycles
(i.e.~adding a further 0.17 ton $\times$ yr exposure) are presented.
The DAMA/LIBRA--phase1 data give evidence for the presence of Dark Matter (DM)
particles in the galactic halo, on the basis of the exploited model independent
DM annual modulation signature by using highly radio-pure NaI(Tl) target, 
at 7.5 $\sigma$ C.L. Including also the first generation DAMA/NaI experiment
(cumulative exposure $1.33$ ton $\times$ yr, corresponding to 14 annual cycles),
the C.L. is 9.3 $\sigma$ and
the modulation amplitude of the {\it single-hit} events in the (2--6) keV energy
interval is: $(0.0112 \pm 0.0012)$ cpd/kg/keV;
the measured phase is $(144\pm 7)$ days 
and the measured period is $(0.998\pm 0.002)$ yr, 
values well in agreement with those expected for DM particles. 
No systematic or side reaction able to mimic 
the exploited DM signature has been found or suggested by anyone over more than a decade.
\end{abstract}

\vspace{5.0mm}


\vspace{2.0mm}


\section{Introduction}

The present DAMA/LIBRA \cite{perflibra,modlibra,modlibra2,bot11,pmts,mu,review,papep,cnc-l,IPP} experiment,
as the former DAMA/NaI \cite{prop,allDM,Nim98,Sist,RNC,ijmd,ijma,epj06,ijma07,chan,wimpele,ldm,allRare,IDM96},
has the main aim to investigate the presence of DM particles in the galactic halo by exploiting
the model independent DM annual modulation signature (originally suggested in Ref.~\cite{Freese}). 
Moreover, the developed highly radio-pure NaI(Tl) target-detectors \cite{perflibra} assure sensitivity to a wide range 
of DM candidates, interaction types and astrophysical scenarios.

As a consequence of the Earth's revolution around the Sun, 
which is moving in the Galaxy with respect to the Local Standard of 
Rest towards the star Vega near
the constellation of Hercules, the Earth should be crossed
by a larger flux of DM particles around $\simeq$ 2 June
and by a smaller one around $\simeq$ 2 December\footnote{Thus, 
the DM annual modulation signature has a different origin and peculiarities
than the seasons on the Earth and than effects
correlated with seasons (consider the expected value of the
phase as well as the other requirements listed below).}.
In the former case the Earth orbital velocity is summed to the one of the
solar system with respect to the Galaxy, while in the latter
the two velocities are subtracted.
The DM annual modulation signature is very distinctive since the effect
induced by DM particles must simultaneously satisfy
all the following requirements: the rate must contain a component
modulated according to a cosine function (1) with
one year period (2) and a phase that peaks roughly 
$\simeq$ 2 June (3); this modulation must only be found in a
well-defined low energy range, where DM particle induced
events can be present (4); it must apply only to those events
in which just one detector of many actually ``fires'' ({\it single-hit}
 events), since the DM particle multi-interaction probability
is negligible (5); the modulation amplitude in the region
of maximal sensitivity must be $\simeq$  7\% for usually adopted
halo distributions (6), but it can be larger in case of some
possible scenarios such as e.g. those in Ref.~\cite{Wei01,Fre04} 
(even up to $\simeq$ 30\%).
Thus this signature is model independent, very effective and, in addition, 
it allows to test a large interval of cross sections and of halo densities.

This DM signature might be mimiced only by systematic effects or side reactions 
able to account for the whole observed modulation amplitude and
to simultaneously satisfy all the requirements given above.
No one is available \cite{perflibra,modlibra,modlibra2,mu,review,Sist,RNC,ijmd}.

\begin{figure}[!ht]
\begin{center}
\includegraphics[width=0.42\textwidth] {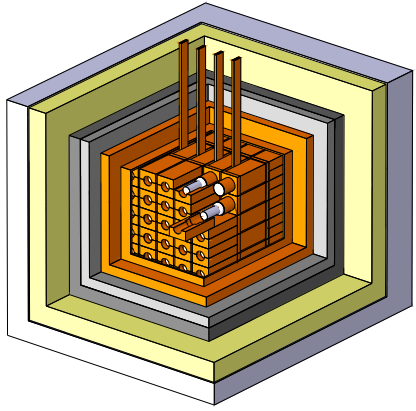}
\end{center}
\vspace{-0.4cm}
\caption{Schematic view of the DAMA/LIBRA apparatus. The 25 highly radiopure NaI(Tl) crystal scintillators
(5-rows by 5-columns matrix), housed in the sealed copper box continuously maintained in
High Purity Nitrogen atmosphere, within low-radioactive passive shield are visible.
Mostly outside the installation, the DAMA/LIBRA apparatus is also almost fully surrounded by 
about 1 m concrete made of the Gran Sasso rock. The copper guides of the calibration system are also shown.
For details see Ref. \cite{perflibra}.}
\label{fg:setup}
\end{figure}

The full description of the DAMA/LIBRA set-up during the phase1 and other related arguments have been discussed in 
details in Ref.~\cite{perflibra,modlibra,modlibra2,review} and references therein.
Here we just remind -- as can be seen in Fig.~\ref{fg:setup} -- that    
the sensitive part of this set-up is made of 25 highly radiopure NaI(Tl) crystal scintillators
(5-rows by 5-columns matrix) having 9.70 kg mass each one.
In each detector two 10 cm long UV light guides (made of Suprasil B quartz) act also as
optical windows on the two end faces of the crystal, and are coupled to two low background
photomultipliers (PMTs) working in coincidence at single photoelectron level. 
The low background 9265-B53/FL and 9302-A/FL PMTs, developed by
EMI-Electron Tubes with dedicated R\&Ds, were used in the phase1; 
for details see Ref.~\cite{perflibra,Nim98,RNC} and references therein.
The detectors are housed in a sealed low-radioactive
copper box installed in the center of a low-radioactive Cu/Pb/Cd-foils/polyethylene/paraffin shield;
moreover, about 1 m concrete (made from the Gran Sasso rock material) almost fully surrounds (mostly
outside the barrack) this passive shield, acting as a further neutron moderator.
A threefold-levels sealing system prevents the detectors to be in contact with the environmental air of the 
underground laboratory \cite{perflibra}. 
The light response of the detectors during phase1 typically ranges 
from 5.5 to 7.5 photoelectrons/keV, depending on the detector. The hardware threshold 
of each PMT is at single photoelectron, while a 
software energy threshold of 2 keV electron equivalent (hereafter keV) is used \cite{perflibra,Nim98}.
Energy calibration with X-rays/$\gamma$ sources are regularly carried out in the
same running condition down to few keV \cite{perflibra}; in particular, 
double coincidences due to internal X-rays from $^{40}$K 
(which is at ppt levels in the crystals) provide (when summing the data over long periods)
a calibration point at 3.2 keV close to the software energy threshold (for details see Ref. \cite{perflibra}).
The DAQ system records both {\it single-hit} events (where just one of the detectors fires) and 
{\it multiple-hit} events (where more than one detector fires) 
up to the MeV region despite the optimization is performed for the lowest one. 
The radiopurity, the procedures and details are discussed in 
Ref.~\cite{perflibra,modlibra,modlibra2,review} and references therein.

The data of the former DAMA/NaI setup (0.29 ton $\times$ yr) and, later, 
those of the first 6 annual cycles of DAMA/LIBRA 
(0.87 ton$\times$yr) 
have already given positive model independent evidence for the presence of DM particles in 
the galactic halo with high confidence level on the basis of the exploited DM annual modulation signature
\cite{modlibra,modlibra2,review,RNC}.

In this paper the final model independent result of DAMA/LIBRA--phase1, obtained by including 
in the analysis also the data collected during the last seventh annual cycle
of operation, is presented. The total exposure of DAMA/LIBRA--phase1 is: 
1.04 ton $\times$ yr; when including also that of the first generation DAMA/NaI experiment it is
$1.33$ ton $\times$ yr, corresponding to 14 annual cycles.

\section{The results}

Table \ref{tb:years} summarizes the information about the seven annual cycles of DAMA/LIBRA--phase1;
the cumulative exposure, considering also the former DAMA/NaI, is also given.

\begin{table}[!hb]
\vspace{-0.2cm}
\caption{Exposures of the 7 annual cycles of DAMA/LIBRA--phase1. Here $\alpha=\langle cos^2\omega 
(t-t_0) \rangle$
is the mean value of the squared cosine, and $\beta=\langle cos \omega (t-t_0) \rangle$
is the mean value of the cosine (the averages are taken over the live time of the data taking
and $t_0=152.5$ day, i.e.~June 2$^{nd}$);
thus, $(\alpha - \beta^2)$ indicates the variance of the cosine
(i.e.~it is 0.5 for a detector being operational evenly throughout the year).
During the first five annual cycles a detector was out of trigger; it was recovered 
in the 2008 upgrade \cite{modlibra2}.}
\begin{center}
\vspace{-0.4cm}
\resizebox{\textwidth}{!}{
\begin{tabular}{|c|c|c|c|c|}
\hline
 & Period & Mass (kg) & Exposure (kg$\times$day)  & $(\alpha - \beta^2)$ \\
\hline
 & & & & \\
 DAMA/LIBRA-1 & Sept. 9, 2003 - July 21, 2004 & 232.8 &  51405  & 0.562 \\ 
   & &        &       & \\ 
 DAMA/LIBRA-2 & July 21, 2004 - Oct. 28, 2005 & 232.8 &  52597  & 0.467 \\ 
   & &        &       & \\ 
 DAMA/LIBRA-3 & Oct. 28, 2005 - July 18, 2006 & 232.8 &  39445  & 0.591 \\ 
   & &        &       & \\ 
 DAMA/LIBRA-4 & July 19, 2006 - July 17, 2007 & 232.8 &  49377  & 0.541 \\ 
   & &        &       & \\ 
 DAMA/LIBRA-5 & July 17, 2007 - Aug. 29, 2008 & 232.8 &  66105  & 0.468 \\ 
   & &        &       & \\ 
 DAMA/LIBRA-6 & Nov. 12, 2008 - Sept. 1, 2009 & 242.5 &  58768  & 0.519 \\ 
   & &        &       & \\ 
 DAMA/LIBRA-7 & Sep.  1, 2009 - Sept. 8, 2010 & 242.5 &  62098  & 0.515 \\
   &        &       & & \\
\hline
 DAMA/LIBRA--phase1 & Sept. 9, 2003 - Sept. 8, 2010 & & 379795 $\simeq$ 1.04 ton$\times$yr & 0.518  \\
\hline
\multicolumn{3}{|l}{DAMA/NaI + DAMA/LIBRA--phase1:} & \multicolumn{2}{c|}{1.33 ton$\times$yr}  \\
\hline
\hline
\end{tabular}
\label{tb:years}}
\vspace{-0.4cm}
\end{center}
\end{table}

The total number of events collected for the energy calibrations during the entire DAMA/LIBRA--phase1 is 
about $9.6 \times 10^7$, while about $3.5 \times 10^6$ events/keV have been collected for 
the evaluation of the acceptance window efficiency for noise rejection near energy 
threshold \cite{perflibra}.

As it can be inferred from Table \ref{tb:years}, the duty cycle of the experiment is high;
the routine calibrations and, in particular, those related with the acceptance windows 
efficiency mainly affect it. The further improvement of the duty cycle 
in the last two annual cycles is mainly due to the 
improved performances of the new transient digitizers and DAQ system installed at fall 2008 
before the start of the sixth annual cycle \cite{modlibra2}. 

The same procedures previously adopted 
\cite{perflibra,modlibra,modlibra2,review} have been exploited also in the analysis of the data of the seventh annual cycle
and several analyses on the model-independent investigation of the DM annual 
modulation signature have been performed.

\begin{figure}[!p]
\begin{center}
\includegraphics[width=0.8\textwidth] {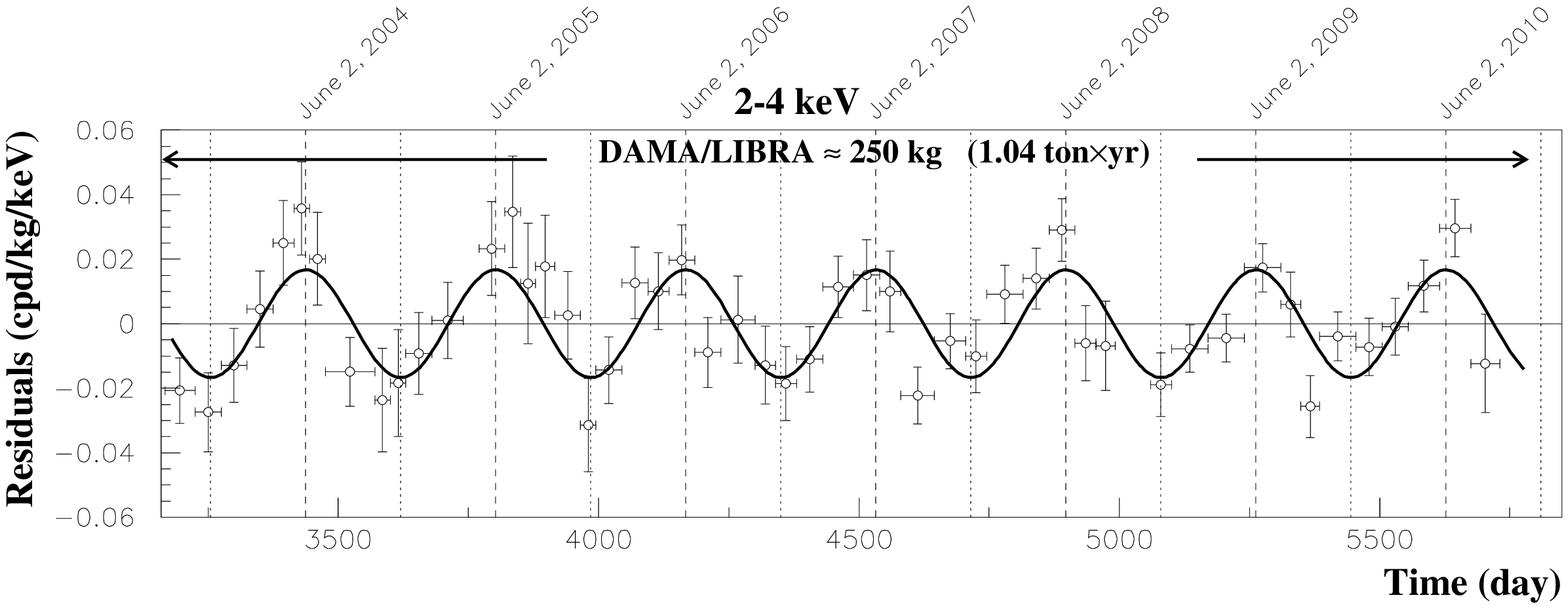}
\vspace{-0.2cm}
\includegraphics[width=0.8\textwidth] {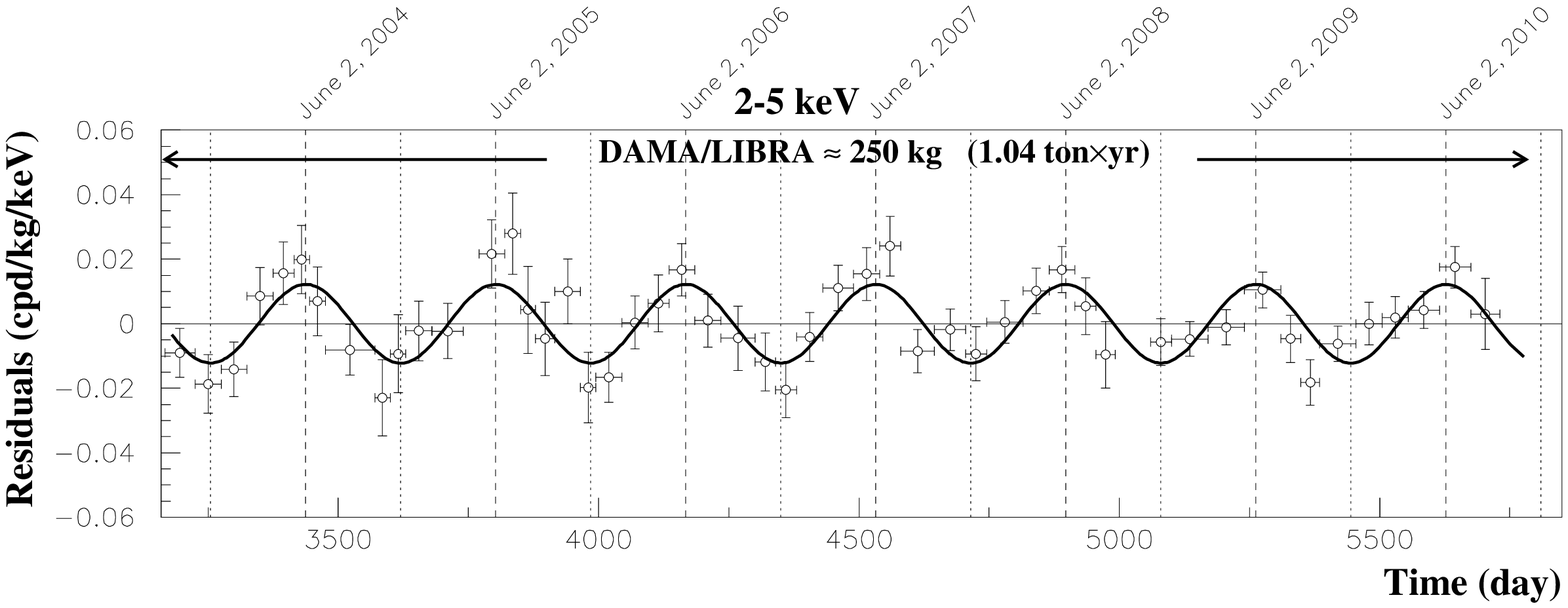} 
\vspace{-0.2cm}
\includegraphics[width=0.8\textwidth] {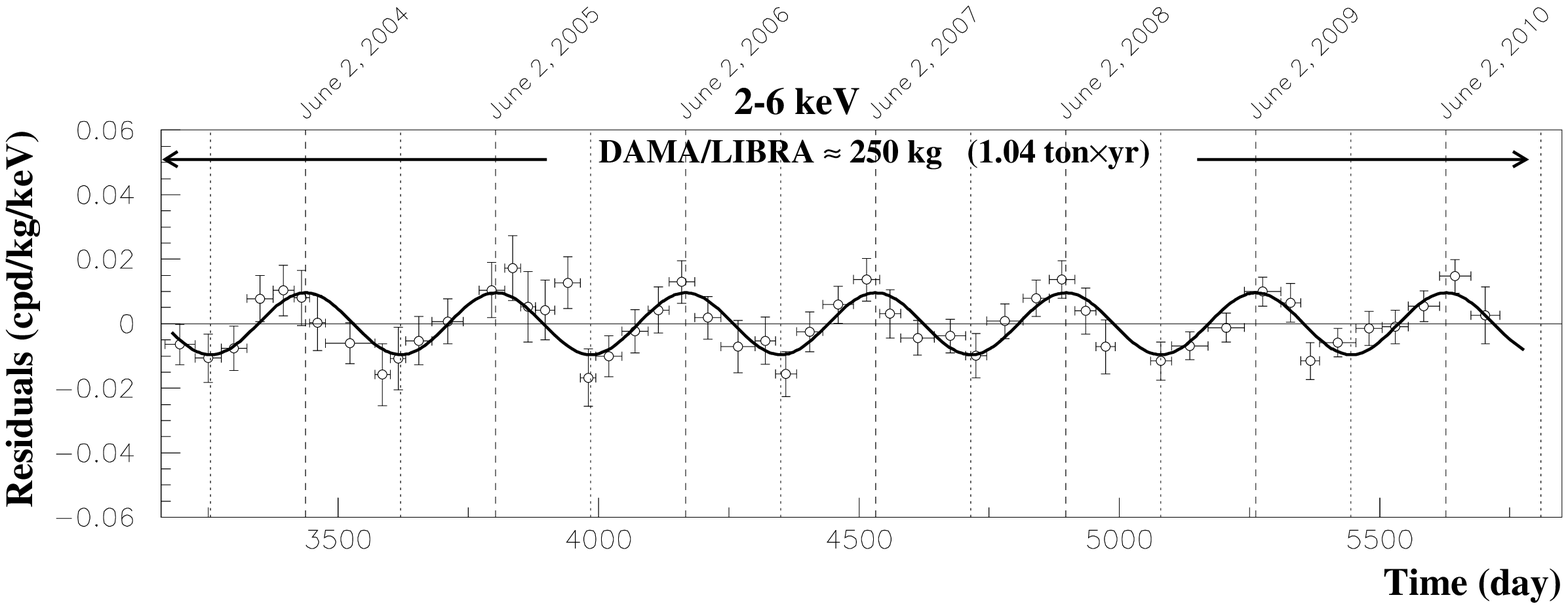}
\end{center}
\vspace{-0.4cm}
\caption{Experimental residual rate of the {\it single-hit} scintillation events
measured by DAMA/LIBRA--phase1 in the (2--4), (2--5) and (2--6) keV energy intervals
as a function of the time. The time scale is maintained the same of the previous DAMA papers
for coherence.
The data points present the experimental errors as vertical bars and the associated
time bin width as horizontal bars.
The superimposed curves are the cosinusoidal functions behaviours $A \cos \omega(t-t_0)$
with a period $T = \frac{2\pi}{\omega} =  1$ yr, a phase $t_0 = 152.5$ day (June 2$^{nd}$) and
modulation amplitudes, $A$, equal to the central values obtained by best fit on the data points of the entire 
DAMA/LIBRA--phase1.
The dashed vertical lines
correspond to the maximum expected for the DM signal (June 2$^{nd}$), while
the dotted vertical lines correspond to the minimum.}
\label{fg:res}
\end{figure}

Fig.~\ref{fg:res} shows the time behaviour of the experimental 
residual rates of the {\it single-hit}  scintillation 
events in the (2--4), (2--5) and (2--6) keV energy intervals for the complete DAMA/LIBRA--phase1. 
The residuals of the DAMA/NaI data (0.29 ton $\times$ yr) are given in Ref.~\cite{modlibra,review,RNC,ijmd}.
We remind that these residual rates are 
calculated from the measured rate of the {\it single-hit} events 
after subtracting the constant part: $<r_{ijk}-flat_{jk}>_{jk}$.
Here $r_{ijk}$ is the rate in the considered $i$-th time interval for the $j$-th detector in the 
$k$-th energy bin, while $flat_{jk}$ is the rate of the $j$-th detector in the $k$-th energy bin 
averaged over the cycles. The average is made on all the detectors ($j$ index) and on all the energy bins ($k$ index) which 
constitute the considered energy interval. The weighted mean of the residuals must 
obviously be zero over one cycle. 
 
The $\chi^2$ test excludes the hypothesis of absence of modulation in the data as shown in Table 
\ref{tb:mod_0}. 
\begin{table}[!ht]
\caption{$\chi^2$ test of absence of modulation in the entire DAMA/LIBRA--phase1 data. The P-values
are also shown. A null modulation amplitude is discarded.}
\begin{center}
\begin{tabular}{|c|rl|}
\hline
 Energy interval & \multicolumn{2}{|c|}{DAMA/LIBRA--phase1} \\
 (keV)           & \multicolumn{2}{|c|}{(7 annual cycles)}  \\
\hline
 2-4 & $\chi^2$/d.o.f. = 111.2/50 & $\rightarrow$ P = 1.5 $\times$ 10$^{-6}$ \\
\hline
 2-5 & $\chi^2$/d.o.f. = 98.5/50  & $\rightarrow$ P = 5.2 $\times 10^{-5}$  \\
\hline
 2-6 & $\chi^2$/d.o.f. = 83.1/50  & $\rightarrow$ P = 2.2 $\times$ 10$^{-3}$  \\
\hline
\hline
\end{tabular}
\end{center}
\label{tb:mod_0}
\end{table}

The {\it single-hit} residual rate of the entire DAMA/LIBRA--phase1 (Fig.~\ref{fg:res}) has been fitted 
with the function: $A \cos \omega(t-t_0)$, considering a
period $T = \frac{2\pi}{\omega} =  1$ yr and a phase $t_0 = 152.5$ day (June 2$^{nd}$) as 
expected by the DM annual modulation signature; this can be repeated including the former 
DAMA/NaI data \cite{RNC} for the cumulative exposure: 1.33 ton $\times$ yr. 
The results of the best fits in the two conditions are summarized in Table \ref{tb:ampff}.
\begin{table}[ht]
\caption{Modulation amplitude, A, obtained by fitting the {\it single-hit} residual rate  
of the entire DAMA/LIBRA--phase1 (Fig.~\ref{fg:res}), and including also 
the former DAMA/NaI data \cite{RNC} 
for a total cumulative exposure of 1.33 ton $\times$ yr. It was obtained by fitting the data with the formula: 
$A \cos \omega(t-t_0)$ with $T = \frac{2\pi}{\omega} =  1$ yr and $t_0 = 152.5$ day (June 2$^{nd}$) as
expected by the DM annual modulation signature. The corresponding $\chi^2$
value of each fit and the confidence level (C.L.) are also reported.}
\begin{center}
\resizebox{\textwidth}{!}{
\begin{tabular}{|c|c|c|}
\hline
 Energy interval & DAMA/LIBRA--phase1 & DAMA/NaI \& DAMA/LIBRA--phase1  \\
 (keV) & (cpd/kg/keV) & (cpd/kg/keV) \\
\hline
 2-4 & A=(0.0167$\pm$0.0022) $\rightarrow$ 7.6 $\sigma$ C.L. & A=(0.0179$\pm$0.0020) $\rightarrow$ 9.0 $\sigma$ C.L. \\
  & $\chi^2$/d.o.f. = 52.3/49 & $\chi^2$/d.o.f. = 87.1/86 \\
\hline
 2-5 & A=(0.0122$\pm$0.0016) $\rightarrow$ 7.6 $\sigma$ C.L. & A=(0.0135$\pm$0.0015) $\rightarrow$ 9.0 $\sigma$ C.L. \\
  & $\chi^2$/d.o.f. = 41.4/49 & $\chi^2$/d.o.f. = 68.2/86 \\
\hline 
 2-6 & A=(0.0096$\pm$0.0013) $\rightarrow$ 7.4 $\sigma$ C.L. & A=(0.0110$\pm$0.0012) $\rightarrow$ 9.2 $\sigma$ C.L. \\
  & $\chi^2$/d.o.f. = 29.3/49 & $\chi^2$/d.o.f. = 70.4/86 \\
\hline
\hline
\end{tabular}}
\end{center}
\label{tb:ampff}
\end{table}

Table \ref{tb:ampfv} shows the results obtained for the entire DAMA/LIBRA--phase1 and including also DAMA/NaI
when the period, and the phase are kept free in the fitting procedure.
The period and the phase are well compatible with expectations for 
a DM annual modulation signal. In particular, the phase is consistent 
with about June $2^{nd}$ and is fully consistent with the value independently determined by Maximum Likelihood 
analysis (see later).
For completeness, we recall that a slight energy dependence of the phase 
could be expected in case of possible contributions of non-thermalized DM components 
to the galactic halo, such as e.g. the SagDEG stream \cite{epj06,Fre05,Gel01} 
and the caustics \cite{caus}.

\begin{table}[!h]
\caption{Modulation amplitude ($A$), period ($T = \frac{2\pi}{\omega}$)
and phase ($t_0$), obtained by fitting, with the formula: 
$A \cos \omega(t-t_0)$, the {\it single-hit} 
residual rate of the entire DAMA/LIBRA--phase1, and including also the former DAMA/NaI data.
The results are well compatible with expectations for a signal in the 
DM annual modulation signature.}
\begin{center}
\begin{tabular}{|rcccc|}
\hline
           & $A$ (cpd/kg/keV)& $T = \frac{2\pi}{\omega}$ (yr) &  $t_0$ (days) & C.L. \\
\hline
\multicolumn{5}{|l|}{DAMA/LIBRA--phase1} \\
 \hspace{1.5cm}  2-4 keV & (0.0178$\pm$0.0022) & (0.996$\pm$0.002) & 134$\pm$7 & 8.1 $\sigma$ \\
 2-5 keV & (0.0127$\pm$0.0016) & (0.996$\pm$0.002) & 137$\pm$8 & 7.9 $\sigma$ \\
 2-6 keV & (0.0097$\pm$0.0013) & (0.998$\pm$0.002) & 144$\pm$8 & 7.5 $\sigma$ \\
     &                     &                   &           &              \\
\multicolumn{5}{|l|}{DAMA/NaI \& DAMA/LIBRA--phase1} \\
 2-4 keV & (0.0190$\pm$0.0020) & (0.996$\pm$0.002) & 134$\pm$6 & 9.5 $\sigma$ \\
 2-5 keV & (0.0140$\pm$0.0015) & (0.996$\pm$0.002) & 140$\pm$6 & 9.3 $\sigma$ \\
 2-6 keV & (0.0112$\pm$0.0012) & (0.998$\pm$0.002) & 144$\pm$7 & 9.3 $\sigma$ \\
\hline
\hline
\end{tabular}
\end{center}
\label{tb:ampfv}
\end{table}

\begin{figure}[!h]
\begin{center}
\vspace{-0.4cm}
\includegraphics[width=10.cm] {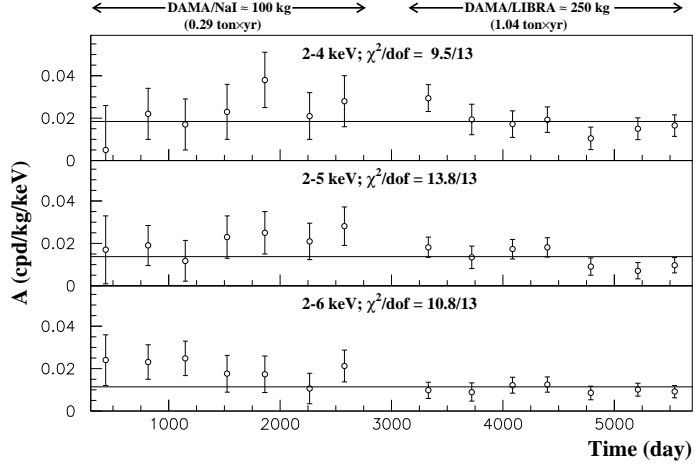}
\end{center}
\vspace{-0.8cm}
\caption{The data points are the modulation amplitudes of each single annual cycle
of DAMA/NaI and DAMA/LIBRA--phase1 experiments. The error bars are the 1$\sigma$ errors.
The same time scale and the same energy intervals as in Fig.~\ref{fg:res} are adopted.
The solid horizontal lines shows the central values obtained by best fit over
the whole data set.
The $\chi^2$ test and the {\it run test} accept the hypothesis at 90\% C.L. that the modulation amplitudes
are normally fluctuating around the best fit values. See text.}
\vspace{-0.3cm}
\label{fg:ampall}
\end{figure}

In Fig.~\ref{fg:ampall} the modulation amplitudes singularly calculated for each 
annual cycle 
of DAMA/NaI and DAMA/LIBRA--phase1 are shown. To test the hypothesis that the 
amplitudes are compatible and normally fluctuating around their mean values
the $\chi^2$ test and the {\it run test} have been performed. 
The mean values in the (2--4) keV, (2--5) keV and (2--6) keV are: 
($0.0185 \pm 0.0020$) cpd/kg/keV,
($0.0138 \pm 0.0015$) cpd/kg/keV and 
($0.0114 \pm 0.0012$) cpd/kg/keV, respectively.
The $\chi^2$ 
value obtained in the (2--4) keV, (2--5) keV and (2--6) keV are 9.5, 13.8 10.8, respectively
over 13 {\it d.o.f.} corresponding to an upper tail probability of 73\%, 39\% and 63\%
for the three energy intervals. We have also performed the {\it run test} 
obtaining a lower tail probabilities of 41\%, 29\% and 23\% for the three 
energy intervals, respectively.
This analysis confirms that the data collected in all the annual cycles with 
DAMA/NaI and DAMA/LIBRA--phase1 are statistically compatible and can be considered together, on the contrary of
the statements in Ref.~\cite{pdp}.

\vspace{0.3cm}

\begin{figure}[!ht]
\centering
\includegraphics[width=5.8cm] {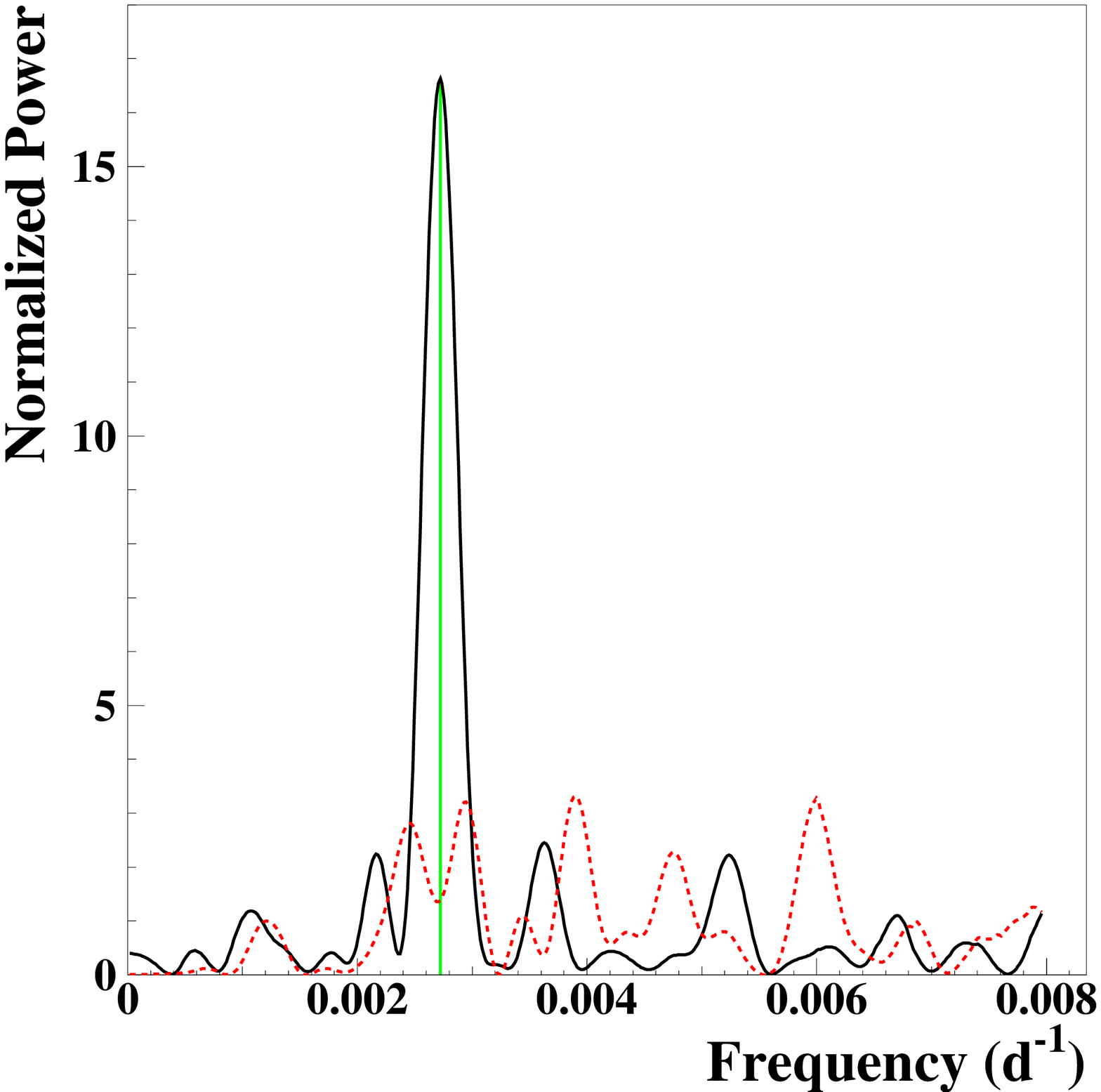}
\includegraphics[width=5.8cm] {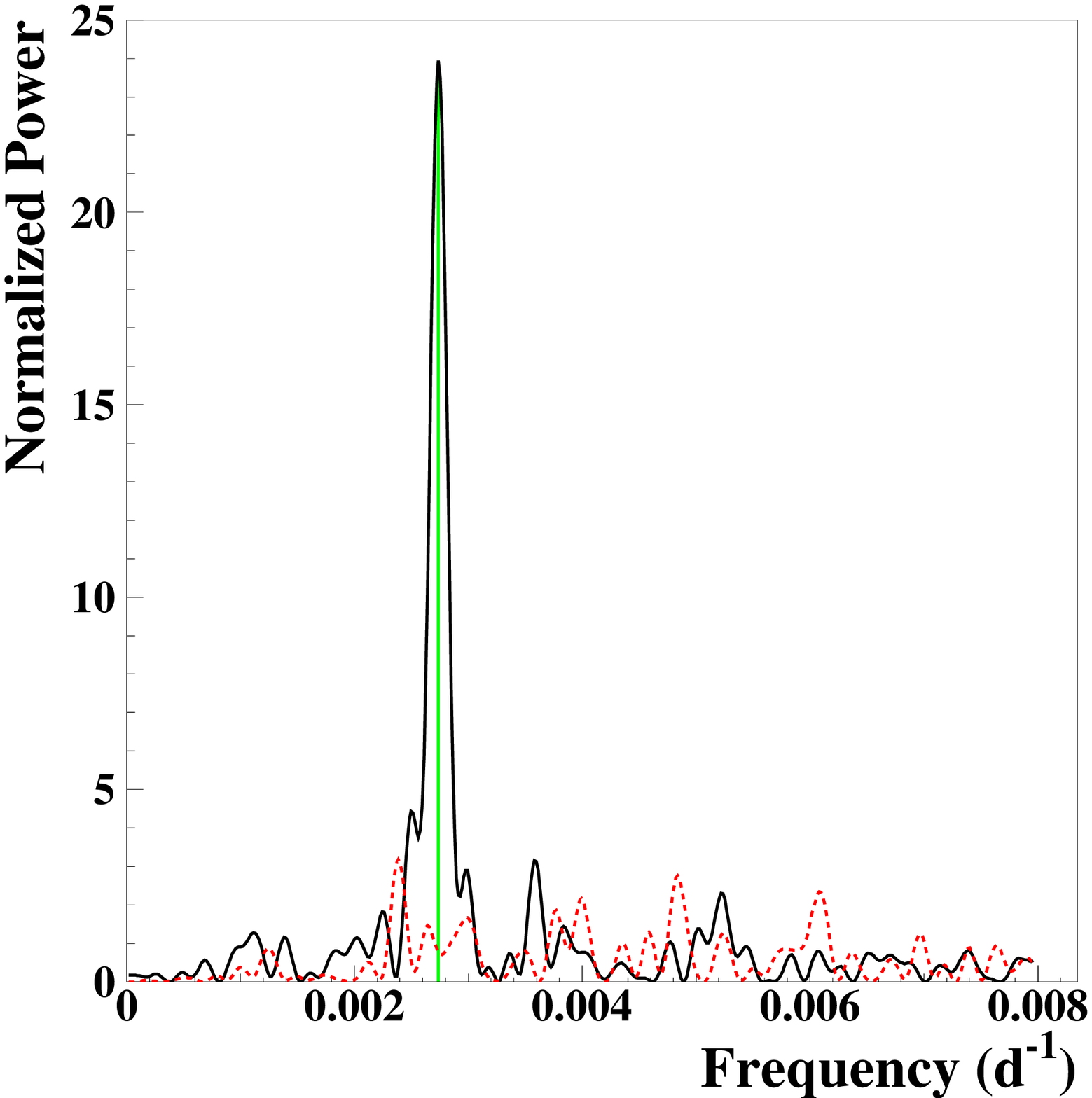}
\vspace{-0.2cm}
\caption{Power spectrum of the measured {\it single-hit} residuals in the (2--6) keV (solid lines) 
and (6--14) keV (dotted lines) energy intervals calculated according to Ref.~\cite{review}, including also 
-- as usual in DAMA analyses --
the treatment of the experimental errors and of the time binning. The data refer to: a) DAMA/LIBRA--phase1 ({\it left}); 
b) DAMA/NaI and DAMA/LIBRA--phase1 ({\it right}). 
As it can be seen, the principal mode present in the (2--6) keV energy interval corresponds to a frequency
of $2.722 \times 10^{-3}$ d$^{-1}$ and $2.737 \times 10^{-3}$ d$^{-1}$ (vertical lines), 
respectively, in the a) and b) case. They correspond to a period 
of $\simeq$ 1 year. A similar peak is not present in the (6--14) keV energy interval.}
\vspace{-0.3cm}
\label{fg:pwr}
\normalsize
\end{figure}

The DAMA/LIBRA--phase1 {\it single-hit} residuals of Fig.~\ref{fg:res} and those of DAMA/NaI 
have also been investigated by a Fourier analysis. The  data analysis procedure 
has been described in details in Ref.~\cite{review}.
A clear peak corresponding to a period of 1 year (see Fig.~\ref{fg:pwr}) is evident
for the (2--6) keV energy interval;
the same analysis in the (6--14) keV energy region shows instead only aliasing 
peaks. Neither other structure at different frequencies has been observed (see also Ref.~\cite{review}).
 
\begin{figure}[!h]
\vspace{-0.7cm}
\begin{center}
\includegraphics[width=4.0cm] {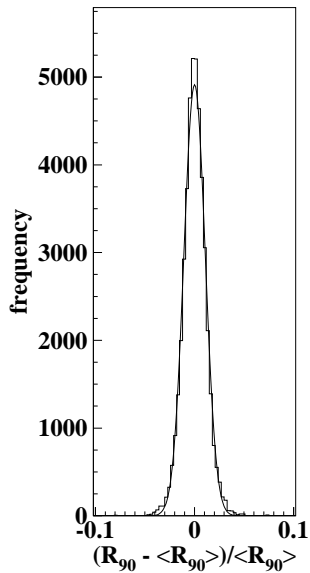}
\end{center}
\vspace{-0.5cm}
\caption{Distribution of the percentage variations
of R$_{90}$ with respect to the mean values for all the detectors in the DAMA/LIBRA--phase1
(histogram); the superimposed curve is a gaussian fit. See text.}
\label{fig_r90}
\end{figure}

\begin{figure}[!t]
\vspace{-0.6cm}
\centering
\includegraphics[width=0.45\textwidth] {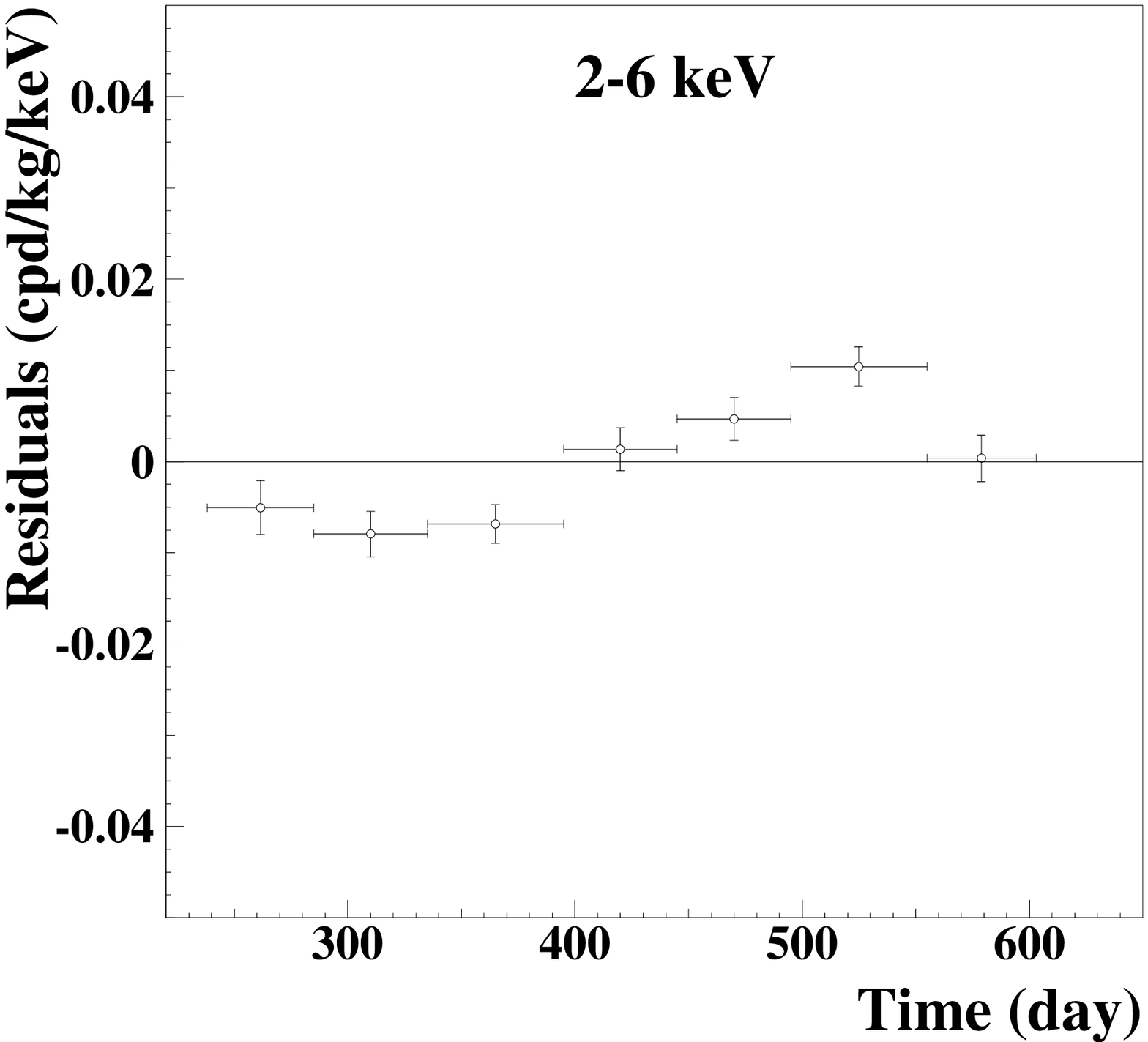}
\includegraphics[width=0.45\textwidth] {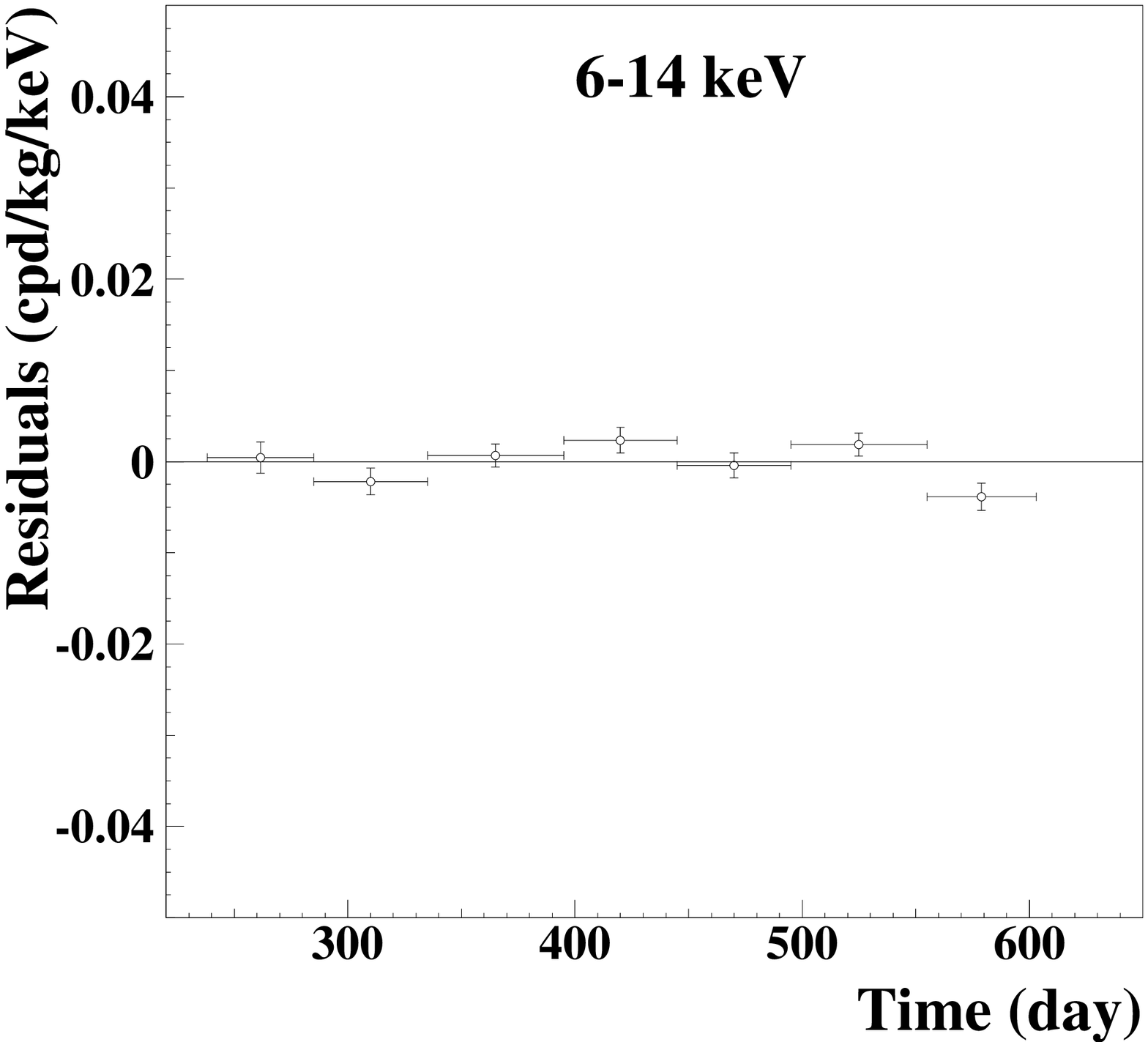}
\vspace{-0.2cm}
\caption{Experimental {\it single-hit} residuals in the (2--6) keV and in the
(6--14) keV energy regions
for the entire DAMA/LIBRA--phase1 data as if they were collected in a
single annual cycle (i.e.~binning in the variable time from the Jan 1st of each annual cycle). 
The data points present the experimental errors 
as vertical bars and the associated
time bin width as horizontal bars. 
The initial time of the figures is taken at August 7$^{th}$.
A clear modulation satisfying all the peculiarities of the 
DM annual modulation signature is present 
in the lowest energy interval with A=(0.0088 $\pm$ 0.0013) cpd/kg/keV,
while it is absent just above: A=(0.00032 $\pm$ 0.00076) cpd/kg/keV.}
\label{fg:res1}
\vspace{-0.4cm}
\end{figure}

\vspace{0.3cm}
Absence of any significant background modulation in the energy spectrum has been verified in
energy regions not of interest for DM\footnote{In fact, the background in the lowest energy region is
essentially due to ``Compton'' electrons, X-rays and/or Auger
electrons, muon induced events, etc., which are strictly correlated
with the events in the higher energy region of the spectrum.
Thus, if a modulation detected in the lowest energy region were due to
a modulation of the background (rather than to a signal),
an equal or larger modulation in the higher energy regions should be present.}.
As an example, the measured rate
integrated above 90 keV, R$_{90}$, as a function 
of the time has been analysed. Fig.~\ref{fig_r90} 
shows the distribution of the percentage variations
of R$_{90}$ with respect to the mean values for all the detectors
in the entire DAMA/LIBRA--phase1 data.
It shows a cumulative gaussian behaviour
with $\sigma$ $\simeq$ 1\%, well accounted by the statistical
spread expected from the used sampling time.
Moreover, fitting the time behaviour of R$_{90}$ including a term
with phase and period as for DM particles, a modulation amplitude compatible with zero
has also been found for all the annual cycles (see Table \ref{tb:r90}).
\begin{table}[!h]
\vspace{-0.3cm}
\caption{
Modulation amplitudes obtained by fitting the time behaviour of R$_{90}$
for the seven annual cycles of DAMA/LIBRA--phase1, including a term with a cosine function
having phase and period as expected for a DM signal. The obtained amplitudes
are compatible with zero, and absolutely incompatible ($\simeq$ 100 $\sigma$)
with modulation amplitudes of tens cpd/kg (see text).}
\begin{center}
\begin{tabular}{|c|c||c|c|}
\hline
 Period & $A_{R_{90}}$ (cpd/kg) & Period & $A_{R_{90}}$ (cpd/kg)   \\
\hline
 DAMA/LIBRA-1 & -(0.05$\pm$0.19) & DAMA/LIBRA-5 & (0.20$\pm$0.18)  \\
 DAMA/LIBRA-2 & -(0.12$\pm$0.19) & DAMA/LIBRA-6 & -(0.20$\pm$0.16) \\
 DAMA/LIBRA-3 & -(0.13$\pm$0.18) & DAMA/LIBRA-7 & -(0.28$\pm$0.18) \\
 DAMA/LIBRA-4 & (0.15$\pm$0.17)  &              &  \\
\hline
\hline
\end{tabular}
\end{center}
\vspace{-0.3cm}
\label{tb:r90}
\end{table}
This also excludes the presence of any background
modulation in the whole energy spectrum at a level much
lower than the effect found in the lowest energy region for the {\it single-hit} events.
In fact, otherwise -- considering the R$_{90}$ mean values --
a modulation amplitude of order of tens cpd/kg would be present for each annual cycle,
that is $\simeq$ 100 $\sigma$ far away from the measured values.
Similar result is obtained when comparing 
the {\it single-hit} residuals in the (2--6) keV with those 
in other energy intervals; for example Fig.~\ref{fg:res1} shows the 
{\it single-hit} residuals in the (2--6) keV and in the
(6--14) keV energy regions for the entire DAMA/LIBRA--phase1 data as if they were collected in a
single annual cycle (i.e.~binning in the variable time from the Jan 1st of each annual cycle). 
It is worth noting that the obtained results account of whatever 
kind of background and, in addition, no background process able to mimic
the DM annual modulation signature (that is able to simultaneously satisfy 
all the peculiarities of the signature and to account for the measured modulation amplitude)
is available (see also discussions e.g. in 
Ref.~\cite{perflibra,modlibra,modlibra2,mu,review,scineghe09,taupnoz,vulca010,canj11,tipp11,replica,replicaA}).

Also in the entire DAMA/LIBRA--phase1 a further relevant investigation has been 
\begin{figure}[!ht]
\begin{center}
\vspace{-0.8cm}
\includegraphics[width=0.9\textwidth] {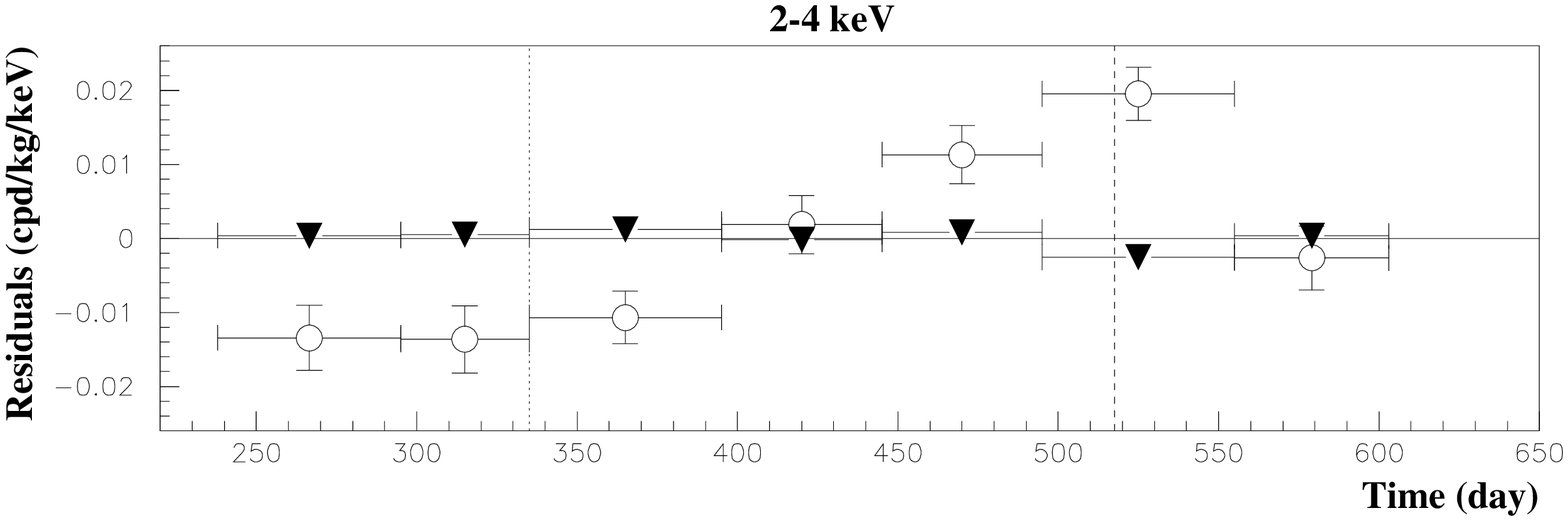}
\vspace{-0.8cm}
\includegraphics[width=0.9\textwidth] {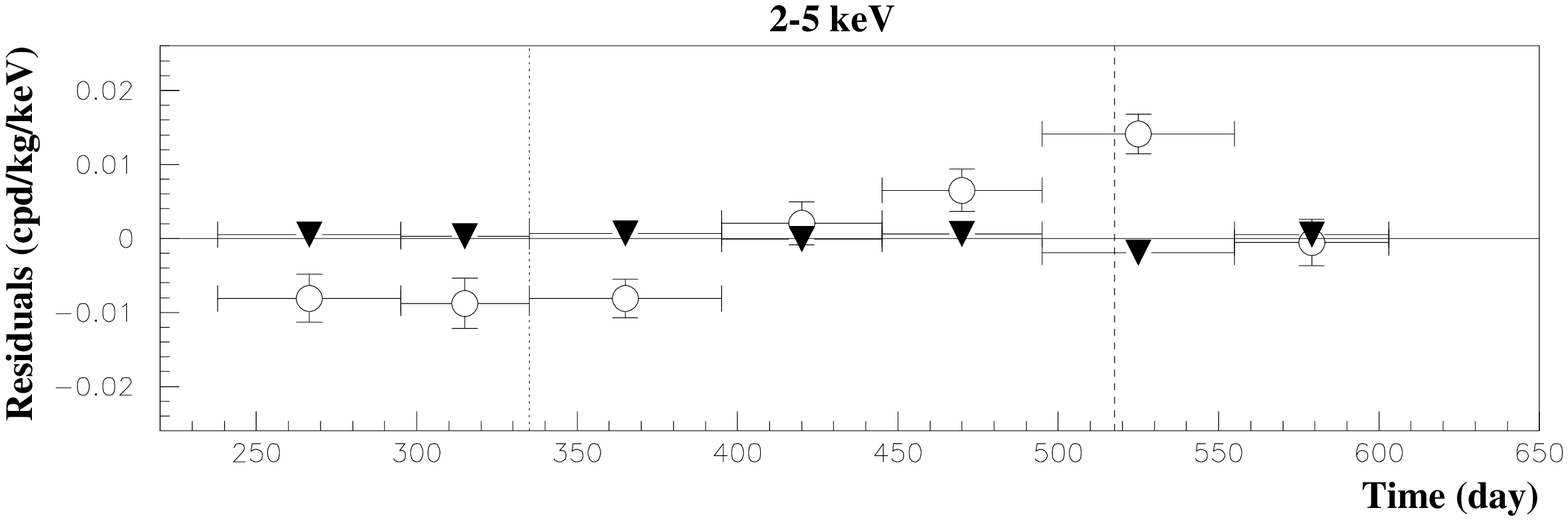}
\vspace{-0.6cm}
\includegraphics[width=0.9\textwidth] {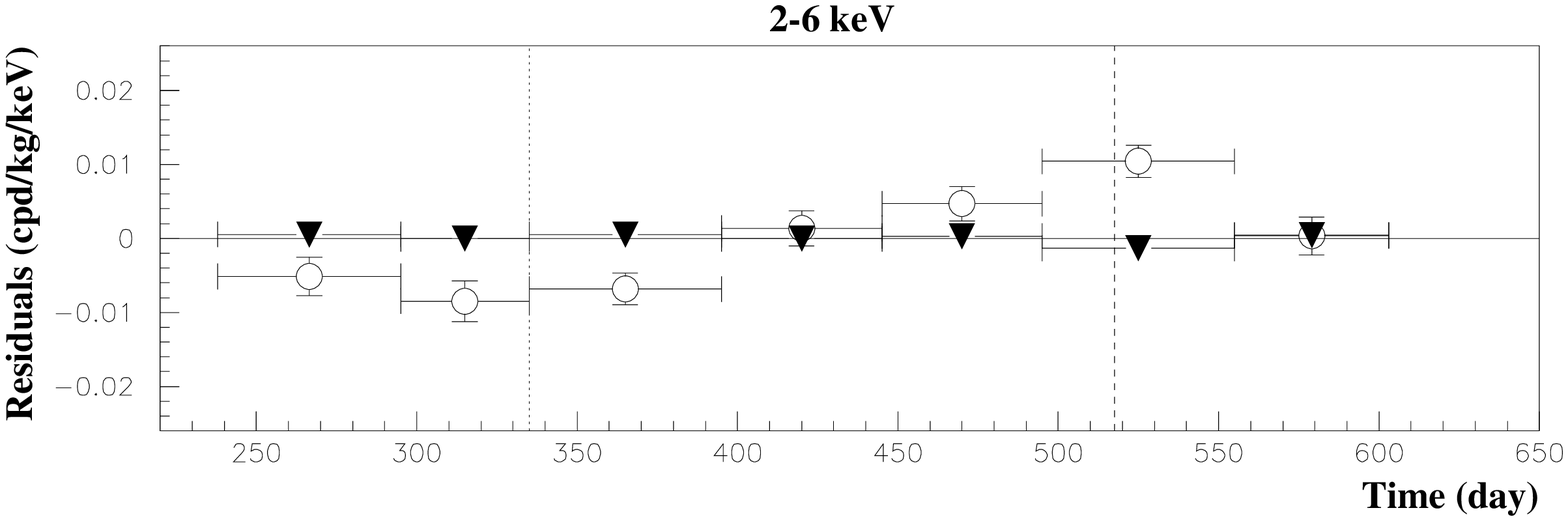}
\end{center}
\vspace{-0.3cm}
\caption{Experimental residual rates over the entire DAMA/LIBRA--phase1 {\it single-hit} events 
(open circles), class of events to which DM events belong, and for {\it multiple-hit} 
events (filled triangles),
class of events to which DM events do not belong.
They have been obtained by considering for each class of events the data as collected in a 
single annual cycle 
and by using in both cases the same identical hardware and the same identical software procedures.
The initial time of the figure is taken on August 7$^{th}$.
The experimental points present the errors as vertical bars 
and the associated time bin width as horizontal 
bars. Analogous results were obtained for the DAMA/NaI data \cite{ijmd}.}
\label{fig_mul}
\end{figure}
performed by 
applying the same hardware and software 
procedures, used to acquire and to analyse the {\it single-hit} residual rate, to the 
{\it multiple-hit} one. In fact, since the 
probability that a DM particle interacts in more than one detector 
is negligible, a DM signal can be present just in the {\it single-hit} residual rate.
Thus, the comparison of the results of the {\it single-hit} events with those of the  {\it 
multiple-hit} ones corresponds practically to compare between them the cases of DM particles beam-on 
and beam-off.
This procedure also allows an additional test of the background behaviour in the same energy interval 
where the positive effect is observed. 
In particular, in Fig.~\ref{fig_mul} the residual rates of the {\it single-hit} events measured over 
the whole DAMA/LIBRA--phase1 annual
cycles are reported, as collected in a single cycle, together with the residual rates 
of the {\it multiple-hit} events, in the considered energy intervals.
While, as already observed, a clear modulation, satisfying all the peculiarities of the DM
annual modulation signature, is present in 
the {\it single-hit} events,
the fitted modulation amplitudes for the {\it multiple-hit}
residual rate are well compatible with zero:
$-(0.0012\pm0.0006)$ cpd/kg/keV,
$-(0.0008\pm0.0005)$ cpd/kg/keV,
and $-(0.0005\pm0.0004)$ cpd/kg/keV
in the energy regions (2--4), (2--5) and (2--6) keV, respectively.
Thus, again evidence of annual modulation with proper features as required by the DM annual 
modulation signature is present in the {\it single-hit} residuals (events class to which the
DM particle induced events belong), while it is absent in the {\it multiple-hit} residual 
rate (event class to which only background events belong).
Similar results were also obtained for the last two annual cycles of the
DAMA/NaI experiment \cite{ijmd}.
Since the same identical hardware and the same identical software procedures have been used to 
analyse the two classes of events, the obtained result offers an additional strong support for the 
presence of a DM particle component in the galactic halo.

\vspace{0.3cm}

As in Ref.~\cite{modlibra,modlibra2,review}, the annual modulation present at low energy can also 
be pointed out by depicting -- as a function of the energy --
the modulation amplitude, $S_{m,k}$, obtained
by maximum likelihood method over the data considering $T=$1 yr and $t_0=$ 152.5 day.
For such purpose the likelihood function of the {\it single-hit} experimental data
in the $k-$th energy bin is defined as: $ {\it\bf L_k}  = {\bf \Pi}_{ij} e^{-\mu_{ijk}}
{\mu_{ijk}^{N_{ijk}} \over N_{ijk}!}$,
where $N_{ijk}$ is the number of events collected in the
$i$-th time interval (hereafter 1 day), by the $j$-th detector and in the
$k$-th energy bin. $N_{ijk}$ follows a Poisson's
distribution with expectation value
$\mu_{ijk} = \left[ b_{jk} + S_{ik} \right] M_j \Delta
t_i \Delta E \epsilon_{jk}$.
The b$_{jk}$ are the background contributions, $M_j$ is the mass of the $j-$th detector,
$\Delta t_i$ is the detector running time during the $i$-th time interval,
$\Delta E$ is the chosen energy bin,
$\epsilon_{jk}$ is the overall efficiency. Moreover, the signal can be written
as $S_{ik} = S_{0,k} + S_{m,k} \cdot \cos\omega(t_i-t_0)$, where $S_{0,k}$ is the constant part of 
the signal and $S_{m,k}$ is the modulation amplitude.
The usual procedure is to minimize the function $y_k=-2ln({\it\bf L_k}) - const$ for each energy bin;
the free parameters of the fit are the $(b_{jk} + S_{0,k})$ contributions and the $S_{m,k}$
parameter. Hereafter, the index $k$ is omitted for simplicity.

\begin{figure}[!ht]
\vspace{-0.5cm}
\begin{center}
\includegraphics[width=\textwidth] {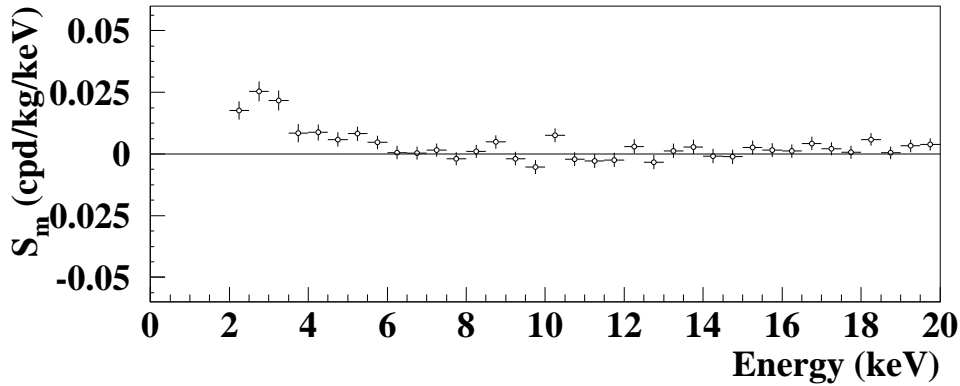}
\end{center}
\vspace{-0.6cm}
\caption{Energy distribution of the $S_{m}$ variable for the
total cumulative exposure 1.33 ton$\times$yr. The energy bin is 0.5 keV.
A clear modulation is present in the lowest energy region,
while $S_{m}$ values compatible with zero are present just above. In fact, the $S_{m}$ values
in the (6--20) keV energy interval have random fluctuations around zero with
$\chi^2$ equal to 35.8 for 28 degrees of freedom (upper tail probability of 15\%).}
\vspace{-0.2cm}
\label{fg:sme}
\end{figure}

\begin{figure}[!thbp]
\vspace{0.2cm}
\begin{center}
\includegraphics[width=0.8\textwidth] {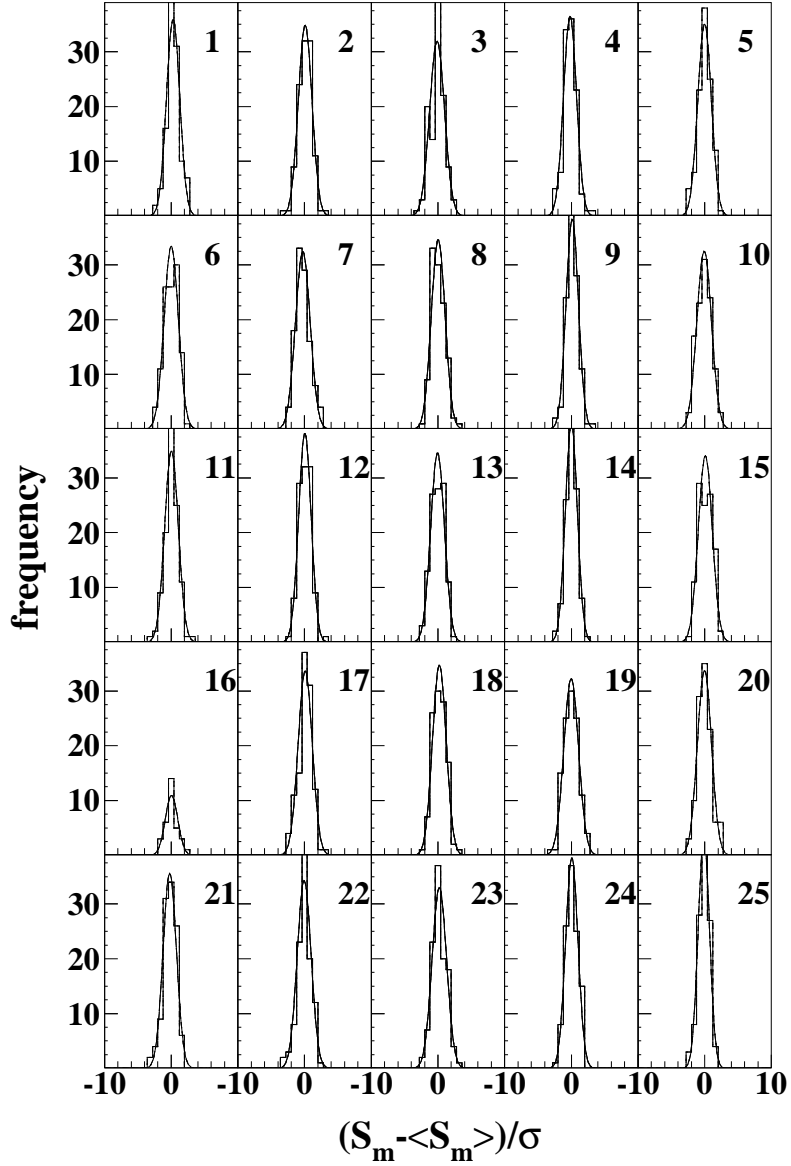}
\end{center}
\vspace{-0.6cm}
\caption{Distributions (histograms) of the variable $\frac {S_m - \langle S_m \rangle}{\sigma}$, where
$\sigma$ are the errors
associated to the $S_m$ values and $\langle S_m \rangle$
are the mean values of the modulation amplitudes averaged over the detectors
and the annual cycles for each considered energy bin (here $\Delta E = 0.25$ keV).
Each panel refers to a single DAMA/LIBRA detector 
(the detector 16 was out of trigger for the first five annual cycles \cite{modlibra2}).
The entries of each histogram are
112 (the 16 energy bins in the (2--6) keV energy interval and the seven DAMA/LIBRA--phase1 annual cycles)
except for detector 16 (32 entries);
the  r.m.s. values are reported in Fig.~\ref{chi2}-{\it bottom}.
The superimposed curves are gaussian fits.}
\label{fg:smcr}
\end{figure}

In Fig.~\ref{fg:sme} the obtained $S_{m}$  are shown 
in each considered energy bin (there $\Delta E = 0.5$ keV) when the data of DAMA/NaI and
DAMA/LIBRA--phase1 are considered.
It can be inferred that positive signal is present in the (2--6) keV energy interval, while $S_{m}$
values compatible with zero are present just above. In fact, the $S_{m}$ values
in the (6--20) keV energy interval have random fluctuations around zero with
$\chi^2$ equal to 35.8 for 28 degrees of freedom (upper tail probability of 15\%).
All this confirms the previous analyses.
As previously done for the other data releases \cite{modlibra,modlibra2,review}, the method also allows the extraction of the the $S_{m}$ 
values for each detector, for each annual cycle and 
for each energy bin. The $S_m$ are expected to follow a normal 
distribution in absence of any systematic effects.
Therefore, the variable $x = \frac {S_m - \langle S_m \rangle}{\sigma}$ has been considered
to verify that the $S_{m}$ are statistically well distributed in all the seven DAMA/LIBRA--phase1 annual cycles,
in all the sixteen energy bins ($\Delta E = 0.25$ keV in the (2--6) keV energy interval)
and in each detector.
Here, $\sigma$ are the errors associated to $S_m$ and $\langle S_m \rangle$
are the mean values of the $S_m$ averaged over the detectors
and the annual cycles for each considered energy bin.
The distributions and their gaussian fits obtained for the detectors are depicted in Fig.~\ref{fg:smcr}.

\vspace{0.2cm}

\begin{figure}[!t]
\vspace{-0.4cm}
\begin{center}
\includegraphics[width=0.6\textwidth] {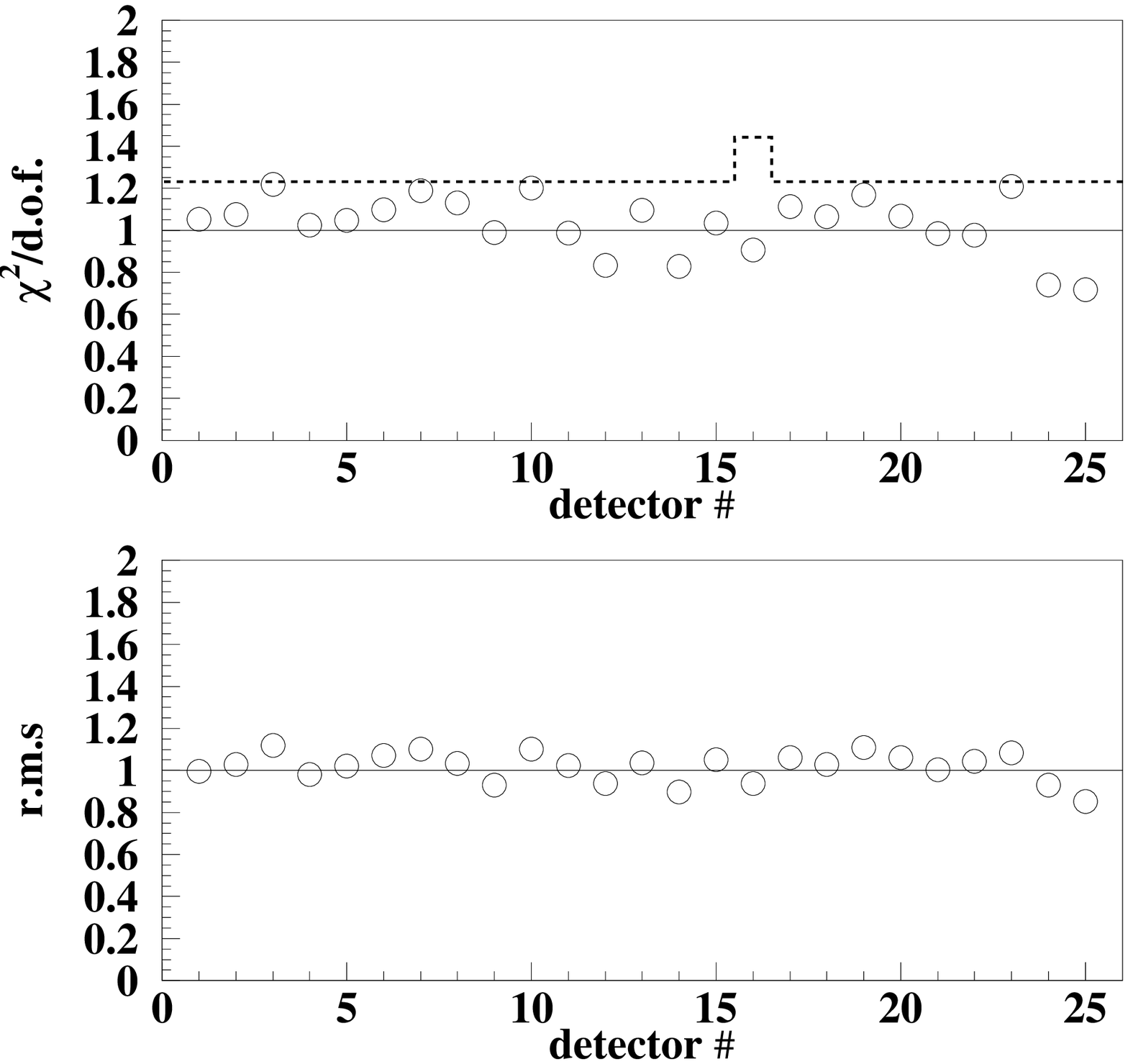}
\end{center}
\vspace{-0.3cm}
\caption{{\it Top :} $\chi^2/d.o.f.$ values of the $S_m$ distributions around their mean value
for each DAMA/LIBRA detector in the (2--6) keV energy interval for the entire DAMA/LIBRA--phase1.
The dotted line corresponds to the upper tail probability of 5\%;
all the $\chi^2/d.o.f.$ values are below this line and, thus, at 95\% C.L.
the observed annual modulation effect is well
distributed in all the detectors. 
{\it Bottom:} standard deviations of the $x$ variable for the DAMA/LIBRA detectors in the phase1.}
\label{chi2}
\end{figure}

Defining $\chi^2 = \Sigma x^2$ -- where the sum is extended over
all the 112 (32 for the detector restored after the 
upgrade in 2008) $x$ values -- $\chi^2/d.o.f.$ values ranging from 0.72 to 1.22 are obtained
(see Fig.~\ref{chi2}{\it --top}); they are all below the 95\% C.L. limit.
Thus the observed annual modulation effect is well
distributed in all the 25 detectors at 95\% C.L. 
The mean value of the 25 $\chi^2/d.o.f.$ is 1.030, slightly larger than 1.
Although this can be still ascribed to statistical fluctuations (see before),
let us ascribe it to a possible systematics. In this case, one would
derive an additional error to the modulation amplitude
measured in the (2--6) keV energy interval:
$\leq 3 \times 10^{-4}$ cpd/kg/keV, if quadratically combining the errors, or
$\leq 2 \times 10^{-5}$ cpd/kg/keV, if linearly combining them.
This possible additional error: $\leq 3\%$ or $\leq 0.2\%$, respectively, on the
DAMA/LIBRA--phase1 modulation amplitude
is an upper limit of possible systematic effects coming from the detector to detector differences.

Among further additional tests, the analysis 
of the modulation amplitudes as a function of the energy separately for
the nine inner detectors and the remaining external ones has been carried out for the 
entire DAMA/LIBRA--phase1. 
The obtained values are fully in agreement; in fact,
the hypothesis that the two sets of modulation amplitudes as a function of the
energy belong to same distribution has been verified by $\chi^2$ test, obtaining:
$\chi^2/d.o.f.$ = 3.9/4 and 8.9/8 for the energy intervals (2--4) and (2--6) keV, 
respectively ($\Delta$E = 0.5 keV). This shows that the
effect is also well shared between inner and outer detectors. 

Let us, finally, release the assumption of a phase $t_0=152.5$ day in the procedure to 
evaluate the modulation amplitudes. In this case the signal can be written as:
\begin{eqnarray}
\label{eqn1} 
S_{ik} & = & S_{0,k} + S_{m,k} \cos\omega(t_i-t_0) + Z_{m,k} \sin\omega(t_i-t_0) \\
       & = & S_{0,k} + Y_{m,k} \cos\omega(t_i-t^*).   \nonumber
\end{eqnarray}
\noindent For signals induced by DM particles one should expect: 
i) $Z_{m,k} \sim 0$ (because of the orthogonality between the cosine and the sine functions); 
ii) $S_{m,k} \simeq Y_{m,k}$; iii) $t^* \simeq t_0=152.5$ day. 
In fact, these conditions hold for most of the dark halo models; however, as mentioned above,
slight differences can be expected in case of possible contributions
from non-thermalized DM components, such as e.g. the SagDEG stream \cite{epj06,Fre05,Gel01} 
and the caustics \cite{caus}.

\vspace{0.4cm}

\begin{figure}[!th]
\begin{center}
\includegraphics[width=0.45\textwidth] {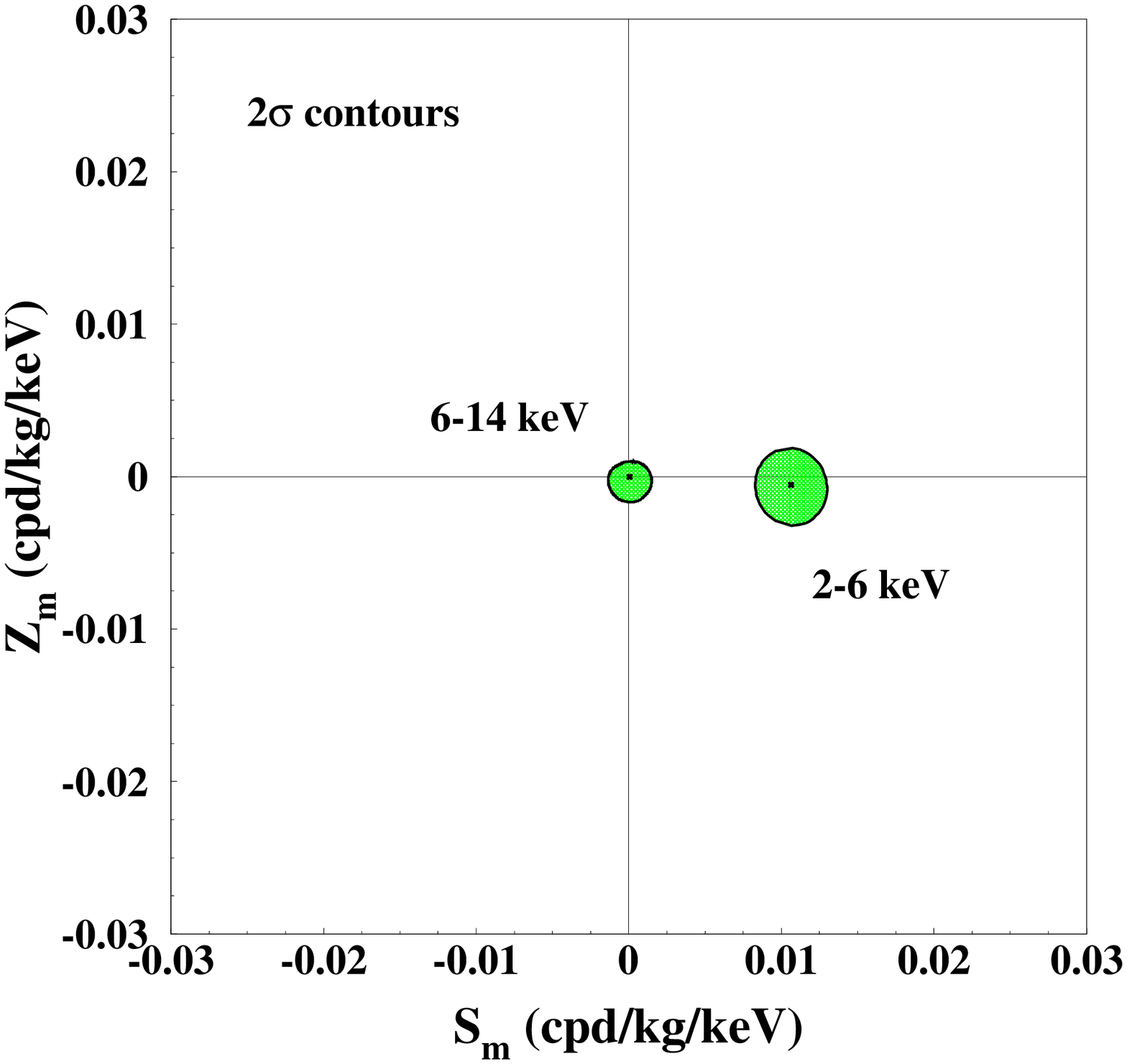}
\includegraphics[width=0.45\textwidth] {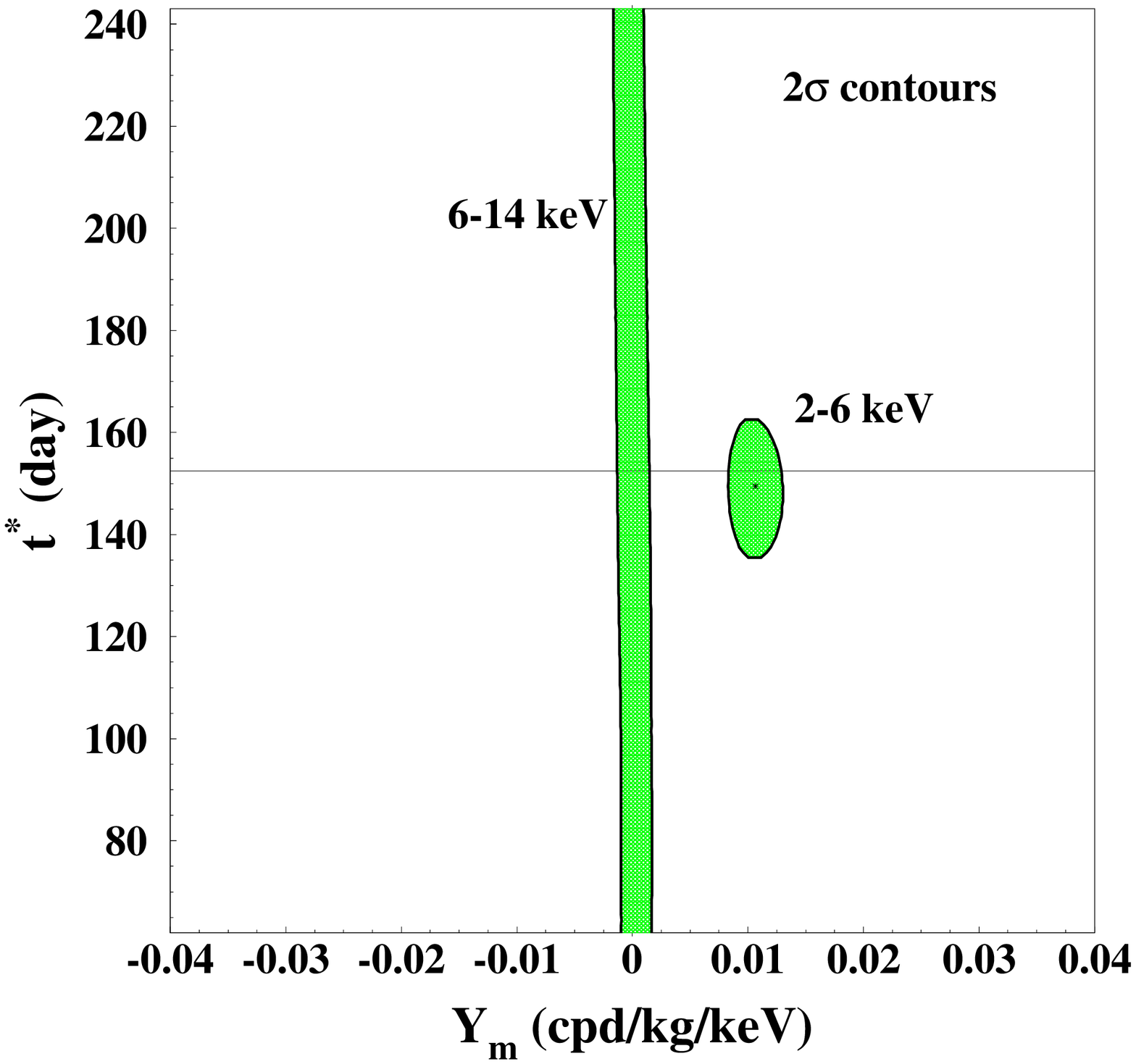}
\end{center}
\vspace{-0.7cm}
\caption{$2\sigma$ contours in the plane $(S_m , Z_m)$ ({\it left})
and in the plane $(Y_m , t^*)$ ({\it right})
for the (2--6) keV and (6--14) keV energy intervals.
The contours have been  
obtained by the maximum likelihood method, considering 
the cumulative exposure of DAMA/NaI and DAMA/LIBRA--phase1.
A modulation amplitude is present in the lower energy intervals 
and the phase agrees with that expected for DM induced signals. See text.}
\label{fg:bid}
\end{figure}

Considering cumulatively the data of DAMA/NaI and DAMA/LIBRA--phase1 (exposure 1.33 ton  $\times$ yr)
the obtained $2\sigma$ contours in the plane $(S_m , Z_m)$ 
for the (2--6) keV and (6--14) keV energy intervals 
are shown in Fig.~\ref{fg:bid}{\it --left} while in 
Fig.~\ref{fg:bid}{\it --right} the obtained $2\sigma$ contours in the plane $(Y_m , t^*)$
are depicted.
The best fit values for the (2--6) and (6--14) keV energy intervals
($1\sigma$ errors) for  S$_m$ versus  Z$_m$ and  $Y_m$ versus $t^*$ are reported in 
Table \ref{tb:bidbf}.

\begin{table}[ht]
\caption{Best fit values for the (2--6) and (6--14) keV energy intervals
($1\sigma$ errors) for  S$_m$ versus  Z$_m$ and  $Y_m$ versus $t^*$,
considering the cumulative exposure of DAMA/NaI and DAMA/LIBRA--phase1.
See also Fig.~\ref{fg:bid}.}
\begin{center}
\resizebox{\textwidth}{!}{
\begin{tabular}{|c||c|c||c|c|}
\hline
E     & S$_m$        & Z$_m$        & $Y_m$        & $t^*$  \\
(keV) & (cpd/kg/keV) & (cpd/kg/keV) & (cpd/kg/keV) & (day) \\
\hline
  2--6 &  (0.0106 $\pm$ 0.0012) & -(0.0006 $\pm$ 0.0012) &  (0.0107 $\pm$ 0.0012) & (149.5 $\pm$ 7.0) \\
 6--14 &  (0.0001 $\pm$ 0.0007) &  (0.0000 $\pm$ 0.0005) &  (0.0001 $\pm$ 0.0008) &  undefined       \\
\hline
\hline
\end{tabular}}
\label{tb:bidbf}
\end{center}
\end{table}

\begin{figure}[!ht]
\begin{center}
\vspace{-0.5cm}
\includegraphics[width=0.85\textwidth] {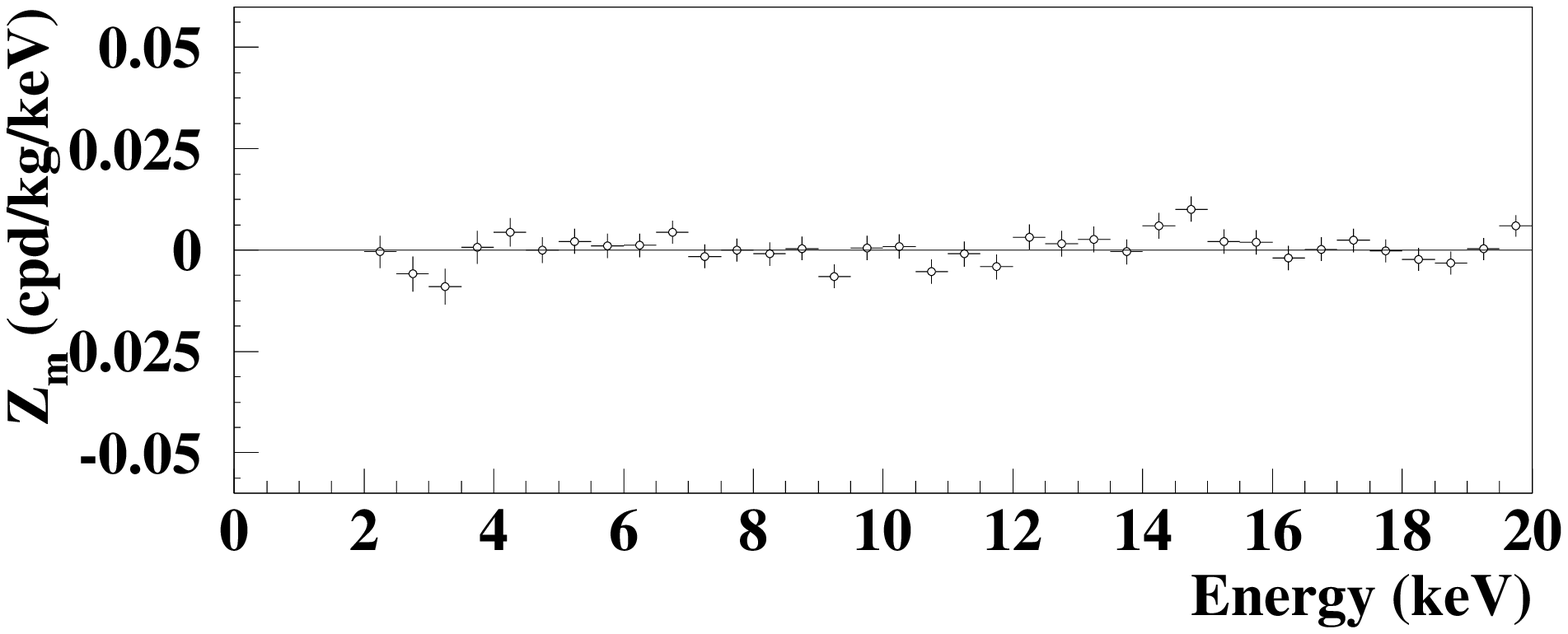}
\end{center}
\vspace{-0.8cm}
\caption{Energy distribution of the $Z_{m}$ variable for the cumulative exposure
of DAMA/NaI and DAMA/LIBRA--phase1, once setting $S_{m}$ in eq. (\ref{eqn1}) to zero.
The energy bin is 0.5 keV.
The $Z_{m}$ values are expected to be zero.
The $\chi^2$ test applied to the data supports the hypothesis that the $Z_{m}$ values are simply 
fluctuating around zero. See text.}
\label{fg:zm}
\end{figure}

Finally, setting $S_{m}$ in eq. (\ref{eqn1}) to zero, 
the $Z_{m}$ values as function of the energy have also been determined
by using the same procedure. The values of $Z_{m}$
as a function of the energy is reported in Fig.~\ref{fg:zm}; they are expected to be zero. 
The $\chi^2$ test 
applied to the data supports the hypothesis that the $Z_{m}$ values are simply 
fluctuating around zero; in fact, for example 
in the (2--14) keV and (2--20) keV energy region the $\chi^2$/d.o.f.
are equal to 23.0/24 and 46.5/36 (probability of 52\% and 11\%), respectively.

\begin{figure}[!ht]
\begin{center}
\includegraphics[width=0.65\textwidth] {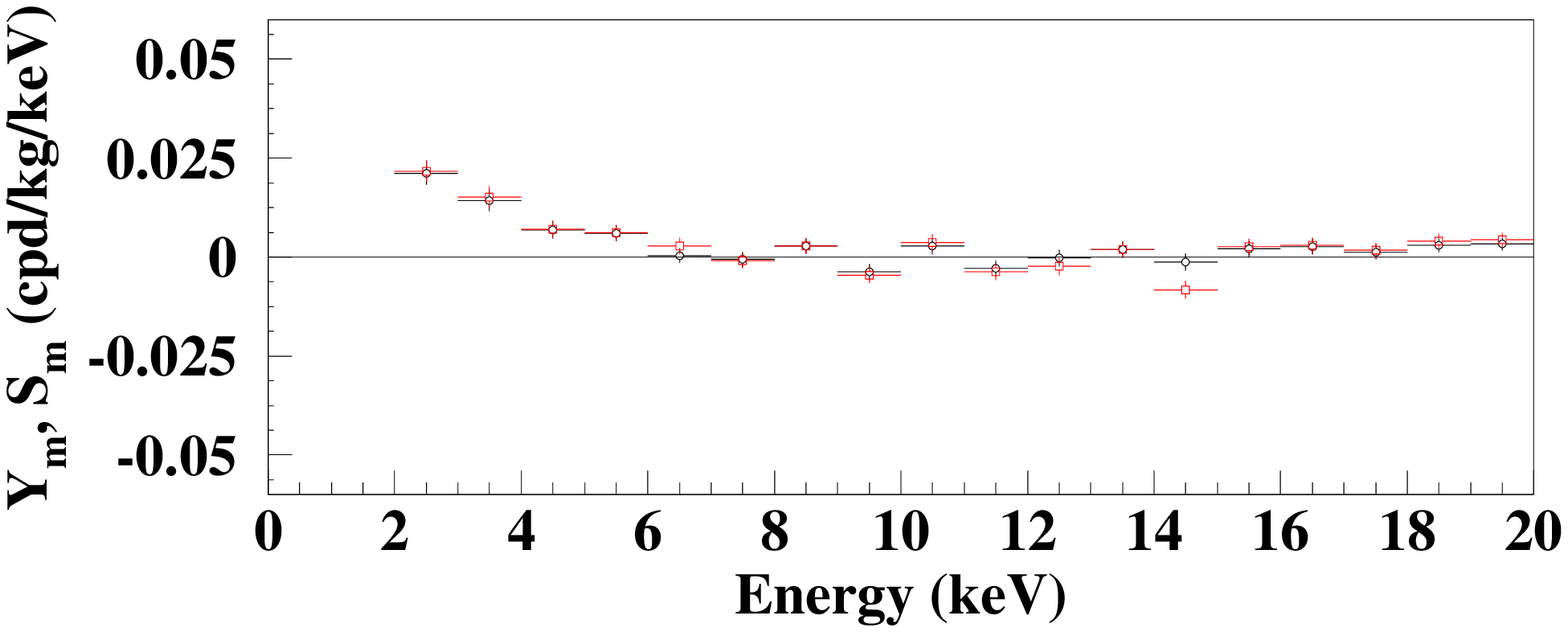}
\includegraphics[width=0.65\textwidth] {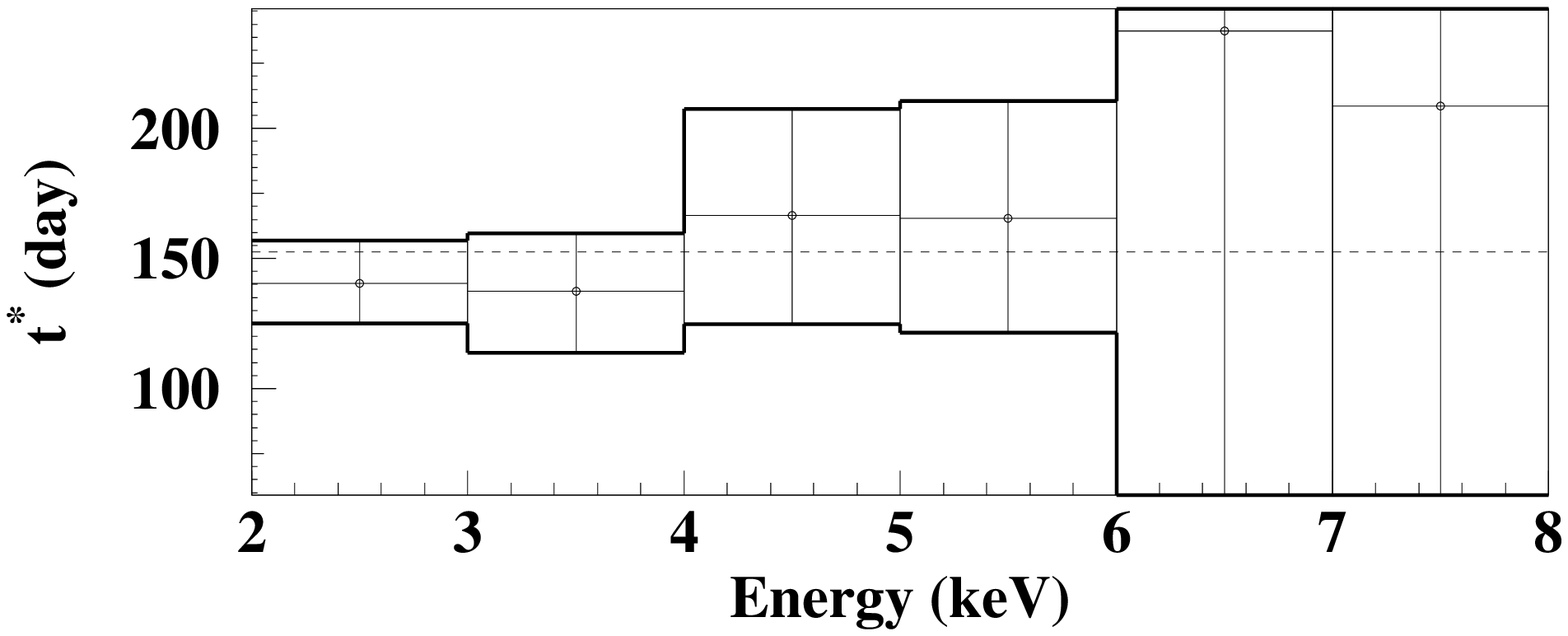}
\end{center}
\vspace{-0.5cm}
\caption{{\em Top:} Energy distributions of the $Y_{m}$ variable (light data points; red colour online)
and of the $S_{m}$ variable (solid data points; black online) for the cumulative exposure
of DAMA/NaI and DAMA/LIBRA--phase1. Here, unlike the data of
Fig.~\ref{fg:sme}, the energy bin is 1 keV.
{\em Bottom:} Energy distribution of the phase $t^*$ for the total exposure; here 
the errors are at $2\sigma$. The vertical scale spans over $\pm$ a quarter of period
around 2 June; other intervals are replica of it.
An annual modulation effect is present in the lower energy intervals 
up to 6 keV and the phase agrees with that expected for DM induced signals.
No modulation is present above 6 keV and thus the phase is undetermined. See text.}
\label{fg:ymts}
\end{figure}

The behaviours of the $Y_{m}$ and of the phase $t^*$ variables 
as function of energy are shown in Fig.~\ref{fg:ymts} 
for the cumulative exposure of DAMA/NaI and DAMA/LIBRA--phase1 (1.33 ton $\times$ yr). 
The $Y_{m}$ are superimposed with the $S_{m}$ values 
with 1 keV energy bin (unlike Fig.~\ref{fg:sme} where the energy bin is 0.5 keV).
As in the previous analyses, an annual modulation effect is present in the lower energy intervals 
and the phase agrees with that expected for DM induced signals.
No modulation is present above 6 keV and the phase is undetermined.

\vspace{0.3cm}

Sometimes naive statements were put forwards as the fact that
in nature several phenomena may show some kind of periodicity.
It is worth noting that the point is whether they might
mimic the annual modulation signature in DAMA/LIBRA (and former DAMA/NaI), i.e.~whether they
might be not only quantitatively able to account for the observed
modulation amplitude but also able to contemporaneously
satisfy all the requirements of the DM annual modulation signature. The same is also for side reactions.
This has already been deeply investigated in Ref.~\cite{perflibra,modlibra,modlibra2} and references
therein;
the arguments and the quantitative conclusions, presented there, also
apply to the entire DAMA/LIBRA--phase1 data. Additional arguments can be found 
in Ref.~\cite{mu,review,scineghe09,taupnoz,vulca010,canj11,tipp11,replica,replicaA}.

\vspace{0.3cm}

In conclusion, the model-independent DAMA results give evidence (at 9.3$\sigma$ C.L. over 14 independent annual cycles)
for the presence of DM particles in the galactic halo.

\vspace{0.5cm}

In order to perform corollary investigation on the nature of the DM particles, model-dependent 
analyses are necessary\footnote{It is worth noting that it does not exist in direct and indirect DM detection experiments 
approaches which can offer such information independently on assumed models.}; thus,
many theoretical and experimental parameters and models are possible
and many hypotheses must also be exploited. In particular, the DAMA model-independent evidence is compatible with a wide 
set of astrophysical, nuclear and particle physics scenarios as also shown in literature.
Moreover, both the negative results and all the possible positive hints, achieved so-far
in the field, are largely compatible with the DAMA model-independent DM annual modulation results in many scenarios considering also the 
existing experimental and theoretical uncertainties; the same holds for indirect approaches. 
For a discussion see e.g. Ref.~\cite{review} and references therein.

\vspace{0.3cm}

Finally, in order to increase the experimental
sensitivity of DAMA/LIBRA and to disentangle -- in the corollary investigation on the candidate particle(s) -- at
least some of the many possible astrophysical, nuclear and particle Physics scenarios \cite{review}, 
the decreasing of the software energy threshold has been pursued. Thus, 
at end of 2010 all the PMTs have been replaced
with new ones having higher quantum efficiency \cite{pmts}; then, the
DAMA/LIBRA--phase2 is started.

\section{Conclusions}

The data of the new DAMA/LIBRA-7 annual cycle have further confirmed a
peculiar annual modulation of the {\it single-hit} events in the (2--6) keV energy region
satisfying all the many requirements of the DM annual modulation signature; the cumulative
exposure by the former DAMA/NaI and DAMA/LIBRA--phase1 is
1.33 ton $\times$ yr. 

In fact, as required by the DM annual modulation signature: 
1) the {\it single-hit} events show a clear cosine-like modulation as expected for the DM signal; 
2) the measured period is equal to $(0.998\pm 0.002)$ yr well compatible with the 1 yr period as expected for the DM signal; 
3) the measured phase $(144\pm 7)$ days is compatible with the roughly $\simeq$ 152.5 days expected for the DM signal; 
4) the modulation is present only in the low energy (2--6) keV interval and not in other higher energy regions, consistently with expectation for the DM signal;
5) the modulation is present only in the {\it single-hit} events, while it is absent in the {\it multiple-hit} ones as expected for the DM signal;
6) the measured modulation amplitude in NaI(Tl) of the {\it single-hit} events in the (2--6) keV energy
   interval is: $(0.0112 \pm 0.0012)$ cpd/kg/keV (9.3 $\sigma$ C.L.).
No systematic or side processes able to simultaneously satisfy all the many
peculiarities of the signature and to account for the whole measured modulation
amplitude is available. 

\vspace{0.3cm}

DAMA/LIBRA is continuously running in its new configuration (named DAMA/LIBRA-phase2) with a lower software energy threshold aiming to improve 
the knowledge on corollary aspects regarding the signal and on second order effects as discussed e.g. in Ref.~\cite{review}.

\section{Acknowledgments} 

It is a pleasure to thank Mr. A. Bussolotti and A. Mattei for their qualified 
technical work.

\end{document}